\documentclass[twocolumn,showpacs,preprintnumbers,amsmath,amssymb]{revtex4}

\usepackage{graphicx}
\usepackage{dcolumn}
\usepackage{bm}

\def\bea{\begin{eqnarray}}
\def\eea{\end{eqnarray}}

\def\pp{\mbox{$p$-$p$}}
\def\ee{\mbox{$e^+$-$e^-$}}
\def\ep{\mbox{$e$-$p$}}
\def\ppbar{\mbox{$p$-$\bar p$}}

\def\auau{\mbox{Au-Au}}

\def\aa{\mbox{A-A}}
\def\nn{\mbox{N-N}}

\def\pt{$p_t$}
\def\yt{$y_t$}

\begin{document} 

\preprint{Version 2.4}

\title{Two-component model of 2D trigger-associated hadron correlations 
on rapidity space $\bf y_{ta} \times y_{tt}$ derived from 1D $\bf p_t$ spectra 
for p-p collisions at $\bf \sqrt{s} = 200$ GeV 
}

\author{Thomas A.\ Trainor}\affiliation{CENPA 354290, University of Washington, Seattle, WA 98195}
\author{Duncan J.\ Prindle}\affiliation{CENPA 354290, University of Washington, Seattle, WA 98195}


\date{\today}

\begin{abstract}
A two-component model (TCM) for single-particle \pt\ spectra describes 200 GeV \pp\  data accurately. Based on that TCM a spectrum hard component was isolated that is related quantitatively to pQCD predictions for jet fragmentation down to low jet energies ($\approx 3$ GeV). We here address jet-related structure in 2D trigger-associated (TA) correlations as a more-detailed method to explore the kinematic limits of low-energy jet production and low-momentum jet fragment structure in \pp\ collisions. We derive a TCM for \pp\ TA correlations that can be used to isolate 2D jet-related structure. Inferred minimum-bias (mainly low-energy) jet-related TA correlations may challenge several major assumptions about jet production in \pp\ (and \aa) collisions. These results should be relevant to \pp\ underlying-event studies and Monte Carlo predictions of multiple parton interactions.
\end{abstract}


\maketitle

 \section{Introduction}

Improved characterization of high-energy \pp\ collisions may extend our understanding of QCD, provide a more accurate reference for heavy ion (\aa) collisions and support LHC searches for physics beyond the Standard Model. While a major effort has been devoted to \pp\ measurements and theoretical analysis some issues remain unclear, specifically the production and manifestations of low-energy jets near the kinematic limits of jet fragment production~\cite{pptheory,fragevo}. Given the steeply-falling spectrum of scattered partons near mid rapidity, {\em minimum-bias} jets are dominated by (mini)jets near an effective spectrum lower bound at 3 GeV, those that are least understood~\cite{anomalous}.
In this study we employ the systematics of 1D \pt\ spectra from 200 GeV \pp\ collisions to derive a two-component model (TCM) for 2D {\em trigger-associated} (TA) correlations. Based on the 2D TCM we propose to extract jet-related {\em hard components} (fragment-pair distributions) from measured TA correlations and determine the kinematic limits for \pp\ minimum-bias jet production.

Extraction of isolated jets from \ee\  and \ep\  collisions has provided accurate determination of in-vacuum jet properties for specific jet energies. Nonperturbative fragmentation functions (FFs) have been measured down to small hadron momenta~\cite{opal,tasso,eeprd} and have been parametrized simply and precisely over a large jet energy range (2 to 200 GeV)~\cite{ppprd}.

Jet systematics in elementary \pp\ and composite \aa\ collisions are less certain. Event-wise jet reconstruction in \pp\ or \ppbar\ collisions, emphasizing high-energy jets, must contend with a nonjet background or {\em underlying event} (UE). Separation of (di)jets from the UE background is typically accomplished with a {\em jet finder} based on a numerical algorithm~\cite{ua1,cdf}. Such algorithms may introduce biases in the inferred jet energy and other jet properties, such as the mean fragment number, fragment momentum distribution and angular distribution. 

For  \aa\ collisions trigger-associated ``dihadron'' 1D azimuth and 2D angular correlations provide a combinatoric jet analysis method for the high-density \aa\ environment~\cite{starhipt}. High-\pt\ {\em trigger} hadrons defined by restrictive \pt\ cuts estimate a jet thrust (parton momentum) axis. Other \pt\ cuts define a class of {\em associated} hadrons based on strong assumptions about jet structure. A combinatoric background is then subtracted based on a physical model designated by the expression ``zero yield at minimum'' (ZYAM). Strong biases in inferred jet yields and angular structure may result from such methods~\cite{tzyam}.

An alternative approach is based on {\em minimum-bias} (MB) 1D  spectrum and 2D two-particle correlation analysis. No a priori assumptions are imposed on jet structure, and no special \pt\ cuts are applied (except an effective low-\pt\ cutoff reflecting a limited detector \pt\ acceptance). Measured event-averaged jet contributions to \pt\ spectra and two-particle correlations emphasize minimum-bias (mainly low-energy) jets. Jet-related spectrum hard components and {\em symmetrized} angular and $p_t \times p_t$ correlations have been measured for \pp\ and \aa\ collisions over a range of collision systems~\cite{ppprd,hardspec,fragevo,porter2,porter3,axialci,anomalous}.

In the present study we extend the MB program. We derive a method to isolate the 2D trigger-associated (TA) correlation structure of minimum-bias jets from \pp\ collisions by analogy with a two-component subtraction method developed for 1D $p_t$ spectrum analysis~\cite{ppprd}. For the 1D analysis a jet-related hard component or {\em fragment distribution} was revealed with quantitative correspondence to pQCD predictions~\cite{fragevo}. In the present study we develop a method to isolate a hard component from {\em asymmetric} (conditional) 2D TA correlations that may be related quantitatively to fragmentation function (FF) data from \ee\ collisions and to a predicted pQCD parton spectrum.

This article is arranged as follows:
Sec.~\ref{anal} summarizes several jet-structure analysis methods.
Sec.~\ref{2comp} reviews two-component models for hadron yields, spectra and two-particle correlations.
Sec.~\ref{trig} presents a 1D two-component model for trigger spectra.
Sec.~\ref{tata} develops a 2D TCM for trigger-associated correlations.
Sec.~\ref{isolate} illustrates how jet-related hard components  can be isolated from TA data.
Sec.~\ref{syserr} estimates systematic uncertainties.
Secs.~\ref{disc} and~\ref{summ} present Discussion and Summary.

 \section{Analysis methods} \label{anal}

In this section we review analysis methods directed to separating jet structure from backgrounds in high-energy nuclear collisions. Jet structure can be inferred by several methods: event-wise 2D jet reconstruction~\cite{opal,cdf}, isolation of nominal jet-related angular correlations using model-dependent $p_t$ cuts and combinatoric-background subtraction~\cite{starhipt} and identification of MB jet fragment distributions within correlations and spectra~\cite{ppprd,hardspec,porter2,porter3}. 
In each case systematic biases may arise from limitations in the method and unjustified assumptions. 

\subsection{Event-wise jet reconstruction}

Event-wise jet reconstruction with a jet-finder algorithm is intended to isolate jet fragments from nonjet hadrons. Because of the complexity of the fluctuating fragmentation cascade jet isolation on $(\eta,\phi)$ is biased at some level. If the jet (parton) energy is the object of measurement systematic bias may be reduced to a relatively low level, since most of the parton energy is carried by a few higher-momentum fragments located near the jet thrust axis (parton momentum).  If the number and properties of low-momentum fragments (the majority) are the object systematic biases may be relatively large, since those hadrons may emerge farther in angle from the thrust axis and be excluded by a jet finder~\cite{pptheory}.

Underlying-event studies are complementary to event-wise jet reconstruction. By assumption UE analysis addresses the nonjet background---what is {\em not} the triggered dijet~\cite{rick,cdf}. Any low-momentum (and large-angle) jet-related hadrons excluded from the dijet by a jet-finder algorithm may be assigned (incorrectly) to the UE. Inferred UE multiplicities, angular correlations and \pt\ distributions could then be misleading~\cite{pptheory}.

\subsection{Combinatoric dihadron analysis and ZYAM}

Jet structure on azimuth inferred from so-called dihadron correlation analysis relies on strong assumptions about background structure and jet fragment distributions. Background estimates invoke measurements of azimuth quadrupole amplitude $v_2$ conventionally associated with ``elliptic flow''~\cite{2004}. But jet structure can contribute strongly to $v_2$ as a ``nonflow'' bias (e.g., the $m=2$ Fourier or quadrupole component of a same-side 2D jet peak narrow on 1D azimuth)~\cite{gluequad,tzyam,multipoles}. And the assumption of zero correlation amplitude at the minimum is inconsistent with substantial same-side and away-side peak overlaps that are expected {\em and observed} for such correlations~\cite{porter2,porter3}. ZYAM background subtraction can then result in underestimated and distorted (severely biased) jet structure~\cite{tzyam,multipoles}. It is further assumed that (a) only ``high-\pt'' hadrons should be associated with pQCD jets and (b) any hadrons below 2 GeV/c must emerge from a thermalized bulk medium in \aa\ collisions~\cite{starbulk,starhipt}.   ``Trigger-associated'' $p_t$ cuts based on such assumptions can further distort and diminish inferred jet structure.

\subsection{p-p collision charge-multiplicity classes}

The charge-multiplicity $n_{ch}$ dependence of spectrum and correlation data from \pp\ collisions provides essential information on collision dynamics and particle production mechanisms. Distinct correlation components are observed to scale quite differently with $n_{ch}$, permitting accurate separation of the individual components without imposing an a priori physical model. Measured systematic properties of the separate components can then be related to candidate collision mechanisms.
Charge multiplicity $n_{ch}$ or its {\em soft component} $n_s$ (defined below) is an analog to \aa\ centrality measures $N_{part}$ and $N_{bin}$ (Glauber parameters). But the relevant \pp\ ``centrality'' parameter may be ``depth'' on nucleon momentum fraction $x$ rather than \pp\ impact parameter $b$~\cite{pptheory}. 
Anticipating future experimental analysis we define seven multiplicity classes ($n_{ch}$ bins, corresponding to fractions of the \pp\ total cross section) over a statistically-significant interval, with bin means $\bar n_{ch}/ \Delta \eta = 1.7, 3.4, 5.5, 7.6, 10, 13.7, 18.7$.

\subsection{Single-particle $\bf y_t$ spectra}

Although transverse momentum $p_t$ is directly measured by particle detectors, {\em transverse rapidity} $y_t \equiv \ln[(m_t + p_t)/m_h]$ (with transverse mass $m_t = \sqrt{p_t^2 + m_h^2}$ and $m_h = m_\pi$ as the default for unidentified hadrons) visualizes important spectrum structure equally well at smaller and larger \yt. In semilog plots on $y_t$ the ``power-law'' data trend arising from the underlying pQCD parton spectrum appears as a straight line at larger $y_t$.

Analysis of the multiplicity dependence of \yt\ spectrum shapes from 200 GeV \pp\ collisions revealed two fixed functional forms---later identified as soft and hard components---scaling approximately as $n_{ch}$ and $n_{ch}^2$~\cite{ppprd}. From subsequent analysis and theory comparisons one component (soft) was associated with projectile nucleon fragmentation proportional to soft multiplicity component $n_s$, and the other component (hard) was associated with dijet production proportional to $n_s^2$~\cite{fragevo}. The combination defines the two-component model for \pp\ spectra.

\subsection{Symmetrized two-particle correlations}  \label{symmcorr}

Symmetrized combinatoric pairs on two-particle momentum space $(\vec p_1,\vec p_2)$ can be factorized with minimal information loss into pair distributions on 2D transverse rapidity space $(y_{t1},y_{t2})$ or $y_t \times y_t$ and on 4D angle space $(\eta_1,\phi_1,\eta_2,\phi_2)$. 
Minimum-bias correlations (no \yt\ cuts) have been studied extensively for \pp~\cite{porter2,porter3} and \auau~\cite{axialci,anomalous} collisions. The quantitative connection between jet-related angular correlations and the $y_t$ spectrum hard component has been demonstrated in Ref.~\cite{jetspec}.

If within some angular acceptance $(\Delta \eta,\Delta \phi)$ the correlation structure is  invariant on pair mean angle (e.g., on $\eta_\Sigma = \eta_1 + \eta_2$), 4D angular correlations can be {\em projected by averaging} onto difference variables (e.g., $\eta_\Delta = \eta_1 - \eta_2$) without loss of information to form {\em angular autocorrelations}~\cite{axialcd,inverse}. The 2D subspace ($\eta_\Delta,\phi_\Delta$) then retains all correlation information and can be visualized. 
The azimuth pair acceptance can be separated into same-side (SS, $|\phi_\Delta| < \pi/2$) and away-side  (AS, $|\phi_\Delta| > \pi/2$) regions.
Features in angular correlations can be modeled by simple functional forms including 1D and 2D Gaussians and sinusoids. Sinusoids $\cos(m\phi_\Delta)$ on azimuth can be characterized as {\em cylindrical multipoles} with pole number $2m$, e.g., dipole, quadrupole and sextupole for $m = 1,~2,~3$. 


Angular correlations on $(\eta_\Delta,\phi_\Delta)$ formed from all combinatoric pairs but excluding ``self'' pairs---particles numerically paired with themselves---are complementary to correlations on transverse rapidity $y_t\times y_t$. The pair distribution is symmetrized about the main diagonal on $y_t \times y_t$. \yt\ acceptance [1,4.5] corresponding to \pt\ interval [0.15,6] GeV/c is consistent with typical detector $p_t$ acceptance and data volumes (statistics limitations at larger \pt). 

Two-particle correlations can be obtained for individual charge-pair types: like-sign (LS), unlike-sign (US), charge-independent (CI = LS $+$ US) and charge-dependent (CD = LS $-$ US). {\em Intra}-jet correlations (same-side jet cone) are dominated by the US combination. {\em Inter}-jet correlations (back-to-back jet pairs) show no charge dependence (CD = 0). Bose-Einstein correlations are observed only for the LS combination. The nonjet azimuth quadrupole conventionally associated with ``elliptic flow'' shows no significant charge dependence.

 \subsection{Conditional trigger-associated correlations}

Minimum-bias {\em conditional} (asymmetric) TA correlations on $(y_{t,assoc},y_{t,trig})$ (with $y_{t,assoc} < y_{t,trig}$  as defined below) are related to, but not equivalent to, symmetrized correlations on $y_t \times y_t$. In either case all nontrivial combinatoric pairs from all events in a given $n_{ch}$ class appear---no particles are rejected by $p_t$ cuts.  But there are quantitative differences in the two structures: Unsymmetrized TA correlations retain additional correlation information, and the TA hard component is more directly comparable to pQCD parton spectrum predictions and measured fragmentation functions.

TA correlations are formed as follows: All events in a given $n_{ch}$ class are separated into ``trigger'' classes ($y_{t,trig}$ or $y_{tt}$ bins) based on the greatest hadron $y_t$ in each event (trigger particle). The single trigger hadron for each event is assumed (with some probability) to be the proxy for a scattered parton. The spectrum for $n_{ch}-1$ {\em associated} hadrons (some may be jet fragments) distributed on $y_{t,assoc}$ or $y_{ta}$ is accumulated for each event in the trigger class, with trigger particles excluded (no self pairs). The resulting 2D TA distribution denoted by $F(y_{ta},y_{tt},n_{ch})$ can be factorized (according to Bayes' theorem) into a trigger spectrum $T(y_{tt},n_{ch})$ (comparable to a pQCD parton spectrum) and a 2D ensemble of distributions $A(y_{ta}:y_{tt},n_{ch})$: associated-hadron spectra conditional on specific trigger $y_{tt}$ values (comparable to an ensemble of fragmentation functions conditional on parton energy).

We employ the two-component model of 1D single-particle $p_t$ spectra derived from 200 GeV \pp\ collisions to define a TCM for 2D TA correlations. By subtracting a 2D soft-component model from TA data we intend to isolate a TA hard component that may be compared quantitatively with pQCD parton spectra and fragmentation function systematics. The method may also be applied to \aa\ collisions to study modified parton fragmentation to jets in more-central collisions.

\section{two-component model} \label{2comp}

The two-component (soft+hard) model~\cite{kn} has been applied to analyses of \pp~\cite{ppprd,pptheory} and \auau~\cite{hardspec,fragevo} collisions. The soft component is attributed to participant-nucleon dissociation 
and the hard component to large-angle-scattered parton fragmentation to jets. The relation of \yt\ spectrum hard-component structure to pQCD theory was established in Ref.~\cite{fragevo}.  
It is also possible to decompose minimum-bias correlation structure according to the TCM. Soft and hard correlation components in more-peripheral \auau\ collisions scale $\propto N_{part}$ (participant nucleons) and $\propto N_{bin}$ (\nn\ binary collisions) respectively~\cite{anomalous,nov2}. Corresponding multiplicity scaling with soft multiplicity $n_s$ for \pp\ collisions (projectile nucleon fragments $\propto n_s$, dijet production $\propto n_s^2$) is discussed in Ref.~\cite{pptheory}.  Identification of the 1D spectrum and 2D correlation hard components with jets in \pp\ and more-peripheral \auau\ collisions is well supported by data systematics and comparisons with pQCD~\cite{ppprd,fragevo,anomalous,jetspec}.

\subsection{Soft and hard event types and multiplicities} \label{tcmmult}

\pp\ collisions can be separated into soft and hard {\em event types}. By definition hard events include at least one minimum-bias jet within the angular acceptance $(\Delta \eta,\Delta \phi)$. Soft events include no jet structure within the acceptance.  Soft and hard event types should be distinguished from soft and hard {\em components} of spectrum and correlation structure from individual events or ensemble averages. 
Each $n_{ch}$ event class with $N_{evt}(n_{ch})$ events includes $N_s$ soft and $N_h$ hard events. Soft $n_s$ and hard $n_h$ multiplicity components averaged over the event ensemble are related by $n_s + n_h = n_{ch}$. Reference~\cite{ppprd} reported that \pp\ hard-component multiplicity $n_h$ scales approximately as $n_h = \alpha\, n_{ch}^2$, implying a preliminary soft-component definition $n_s = n_{ch} -  \alpha\, n_{ch}^2$. A more accurate hard-component trend is $n_h = \alpha\, n_{s}^2$ with $\alpha \approx 0.006$ (for acceptance $\Delta \eta = 1$).

The observed mean dijet number $n_j$ within acceptance $\Delta \eta$ varies with soft-component multiplicity $n_s$ as  $n_j = 0.015\, \Delta \eta\, (n_s / 2.5\Delta \eta)^2$ scaled from non-single-diffractive (NSD) \pp\ collisions~\cite{ppprd}.  Poisson probabilities for soft and hard events are then $P_s(n_s) = N_s/N_{evt} = \exp(-n_j)$ and $P_h(n_s) = N_h/N_{evt} = 1- P_s(n_s)$. For soft events $n_s'' = n_{ch}$ and for hard events $n_s' + n_h' = n_{ch}$, defining those symbols. The several $n_{ch}$ components then satisfy the following relations: (i) $n_{ch} = n_s + n_h = P_s n''_{s} + P_h (n_s' + n_h')$, (ii) $n_s = P_s n_{ch} + P_h n_s'$ and $n_h = P_h n_h' = n_j \bar n_{ch,j}$, (iii)
$n_s' = n_s - (P_s/P_h) n_h = n_s - P_s n_h' \approx n_s - P_s \bar n_{ch,j}$, where $\bar n_{ch,j}$ is the ensemble-mean dijet fragment multiplicity within $\Delta \eta$. For smaller event multiplicities $P_h \approx n_j \propto n_s^2$.

\subsection{p-p single-particle $\bf y_t$ spectra} \label{ppspecc}

\begin{figure}[h]
  \includegraphics[width=1.65in,height=1.6in]{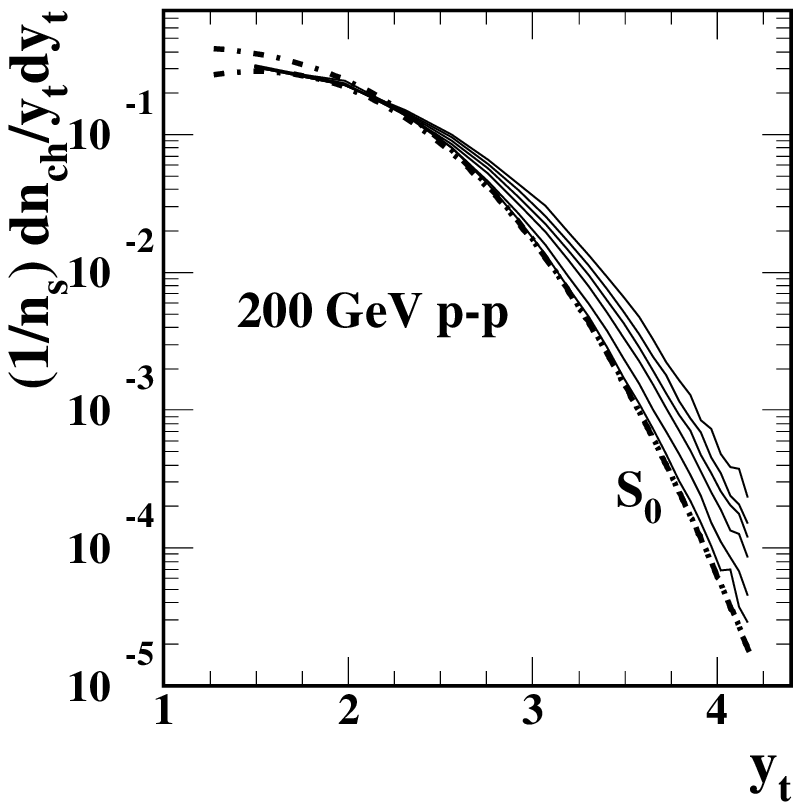}
  \includegraphics[width=1.65in,height=1.6in]{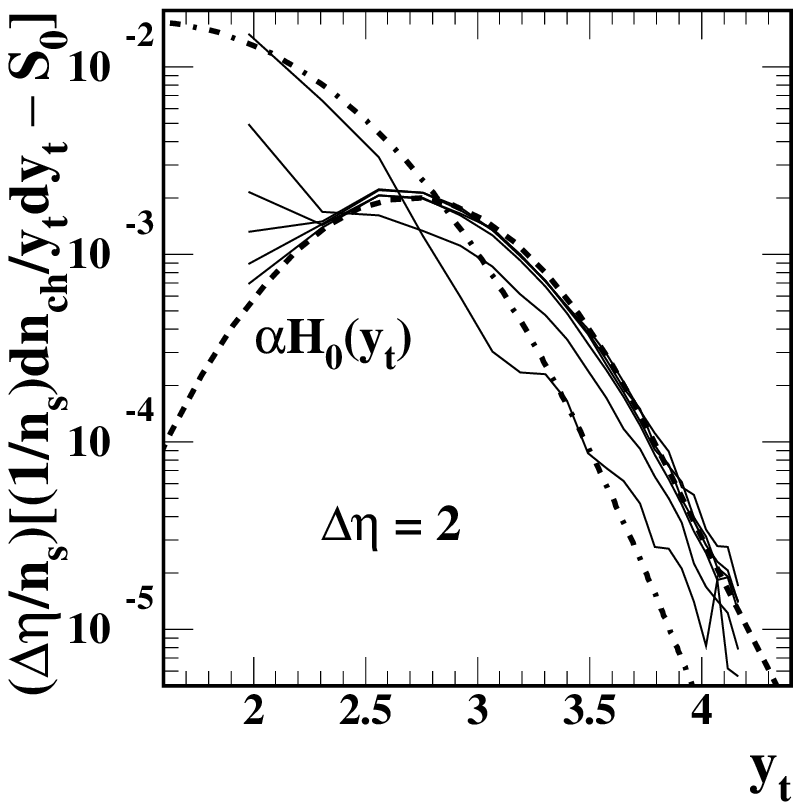}
\caption{\label{ppspectra}
Left: Normalized single-particle spectra (thin solid curves) from seven multiplicity $n_{ch}$ classes of 200 GeV \pp\ collisions. The dash-dotted curves are inferred soft component $S_0$ for uncorrected (lower) and efficiency-corrected (upper) spectra.
Right: Hard components (thin solid curves) extracted from spectra in the left panel according to the vertical axis label. The bold solid curve is the jet-related hard-component model function $H_0$ with coefficient $\alpha = 0.006$. The dash-dotted curve is soft component $S_0$ scaled by $\Delta \eta / n_{s}$  for the highest multiplicity class (7).
 }  
\end{figure}

Distinct soft and hard components of 1D $y_t$ spectra from \pp\ collisions are observed to have fixed forms on \yt, but their relative amplitudes vary with $n_{ch}$ (or $n_s$). The two-component model of $y_t$ spectra $\rho(y_t,n_{ch}) = \rho_0( n_{ch}) F_0(y_t, n_{ch})$ from \pp\ collisions conditional on measured $ n_{ch}$ integrated within some angular acceptance $(\Delta \eta,\Delta \phi)$ is described by~\cite{ppprd}
\bea \label{ppspec}
\rho_0( n_{ch}) F_0(y_t, n_{ch}) &=& S(y_t,n_{ch}) + H(y_t,n_{ch}) \\ \nonumber
&=&  \rho_s( n_{ch}) S_0(y_t)  +  \rho_{h}( n_{ch}) H_0(y_t),~~~
\eea
where $\rho_0 = n_{ch} / \Delta \eta \Delta \phi$ is the $y_t$-integral single-particle (SP) angular density, and $\rho_s = n_s / \Delta \eta \Delta \phi$ and $\rho_h = n_h / \Delta \eta \Delta \phi$ are corresponding soft and hard charge angular densities. 
The inferred spectrum soft and hard model components [unit integral $S_0(y_t)$ and $H_0(y_t)$ functions, see App.~\ref{tcmmodelfunc}] are independent of $n_{ch}$.  Soft-component model $S_0(y_t)$ is defined as the limiting form of the normalized spectra appearing in Fig.~\ref{ppspectra} (left) as $n_s \rightarrow 0$. Hard-component model $H_0(y_t)$ represents data hard components $H(y_{t},n_{ch})$ obtained from spectrum data by subtracting the soft-component model and is predicted by measured parton fragmentation functions (from \pp\ collisions) folded with a minimum-bias pQCD parton (dijet) spectrum integrating to $\sigma_{dijet} \approx 2.5$ mb~\cite{fragevo}.

Figure~\ref{ppspectra} (left) shows $y_t$ spectra for several multiplicity classes (thin solid curves). Each spectrum is normalized by a corresponding soft multiplicity $n_s$ inferred iteratively (Sec.~\ref{tcmmult}). The bold dash-dotted curves in the left panel represent soft component $S_0(y_t)$. The plotted spectra are uncorrected for detector inefficiencies. The $y_t$-averaged inefficiency cancels in the normalization ratio. A $y_t$-dependent inefficiency function deviating from unity at lower $y_t$ and included in the $S_0$ model to relate corrected and uncorrected spectra is indicated by the ratio of the two dash-dotted curves.

Figure~\ref{ppspectra} (right) shows the {\em measured} hard components $H(y_t,n_{ch})$ (thin solid curves) inferred by subtracting the soft-component model $S_0(y_t)$ from each normalized data spectrum in the left panel and normalized by soft density $n_s / \Delta \eta$ with acceptance $\Delta \eta = 2$. Data for the higher $n_{ch}$ classes coincide with TCM function $H_0(y_t)$ (bold dashed curve) within data uncertainties except for the lowest $y_t$ points, but the data for two $n_{ch}$ classes fall significantly below the majority. $H(y_t)$ for the first $n_{ch}$ bin is reduced by factor 3 (and the next bin by a smaller factor) from the principal trend, as noted in Ref.~\cite{ppprd}. 
The observed systematic deviations are included in the TA TCM defined below.
The dashed curve is $\alpha H_0(y_t)$.  Hard-component model $H_0(y_t)$ is a Gaussian {\em with power-law tail} as introduced in Ref.~\cite{hardspec} and described in App.~\ref{tcmmodelfunc}. A preliminary value $\alpha = 0.005 \pm 0.0015$ was reported in Ref.~\cite{ppprd}. The relation of $\alpha$ to dijet production is discussed below.
The TCM for 1D \yt\ spectra reveals minimum-bias jet structure (hard component) in quantitative agreement with pQCD predictions~\cite{ppprd,hardspec,fragevo,jetspec}.  

The dash-dotted curve in the right panel is soft component $S_0$ scaled according to the $y$-axis label for the highest multiplicity class. Direct comparison with the ensemble of hard components indicates that the TCM hard component is accurately determined down to $y_t \approx 2$ by subtraction of model $S_0$, especially for larger $n_{ch}$.

Measured SP spectra integrated over some angular acceptance can also be decomposed in terms of soft and hard event types as
\bea \label{spevtype}
\frac{d n_{ch}}{y_t dy_t} &=& P_s(n_{ch}) S_s(y_t,n_{ch}) \\ \nonumber
&+& P_h(n_{ch}) [S_h(y_t,n_{ch}) + H_h(y_t,n_{ch})]
\eea
with corresponding TCM model elements $S_s \rightarrow n_{ch} S_0$, $S_h \rightarrow n_s' S_0$ and $H_h \rightarrow n_h' H_0$, and with $n_h' \approx \bar n_{ch,j}$ for smaller $n_{ch}$.

\subsection{p-p minimum-bias dijet production}

Equation~(\ref{ppspec}) integrated over some angular acceptance $(\Delta \eta,\Delta \phi)$ becomes
\bea \label{ppspec2}
\frac{d n_{ch}}{y_t dy_t} &=& n_s S_0(y_t) + n_h H_0(y_t).
\eea
The measured data trend in Fig.~\ref{ppspectra} (right) implies $n_h/n_s = \alpha\, n_s/\Delta \eta$ or 
\bea \label{freq}
\frac{n_h}{\Delta \eta} &=& 0.006 \pm 0.001 \left( \frac{n_s}{\Delta \eta}\right)^2 \\ \nonumber
&\equiv& f(n_{ch}) \bar n_{ch,j}(\Delta \eta),
\eea
where the second line represents the jet hypothesis from Ref.~\cite{ppprd} and defines dijet frequency $f = dn_j/ d\eta$ (dijet number per event per unit $\eta$) with mean dijet fragment multiplicity $\bar n_{ch,j}$.

That result can be compared with a pQCD prediction of the dijet frequency for non-single-diffractive (NSD) \pp\ collisions. From the trend in Eq.~(\ref{freq}) we have
\bea
f(n_{ch}) &=& f_{NSD} \frac{n_{s}^2}{n_{s,NSD}^2}  =\frac{f_{NSD}}{(n_{s,NSD}/\Delta \eta)^2} \left(\frac{n_s}{\Delta \eta}\right)^2 \\ \nonumber
f_{NSD} &=& \frac{0.006 \times 2.5^2}{\bar n_{ch,j}(\Delta \eta)} \approx 0.015,
\eea
with $n_s / \Delta \eta \approx 2.5$ for NSD \pp\ collisions and $\bar n_{ch,j}(\Delta \eta) \approx 2.5$~\cite{eeprd}. The value  $f_{NSD} \approx 0.015$ inferred from the \pp\ spectrum $n_{ch}$ dependence is consistent with a pQCD prediction 0.015 based on a parton spectrum cutoff near 3 GeV corresponding to dijet total cross section 2.5 mb~\cite{fragevo}. Thus, we obtain from a TCM analysis of 1D \yt\ spectra  the dijet $\eta$ density $f = dn_j/d\eta$ as a function of $n_s$.

The minimum-bias dijet rate determined directly  from \pp\ \yt\ spectra can be contrasted with Monte Carlo parametrizations~\cite{hijing,pythia,herwig,herwig2}. Dijet production in \aa\ collisions modeled by \textsc{hijing} (based on the \textsc{pythia} model of \pp\ collisions) is determined by an effective \pp\ dijet total cross section $\sigma_{dijet} \approx 25$ mb~\cite{liwang}, ten times the value inferred from measured \yt\ spectra~\cite{fragevo}. 
\textsc{hijing} predictions for \auau\ hadron production disagree strongly with data trends, prompting some to reject the TCM for heavy ion collisions~\cite{review}. 
Ironically, while the \textsc{hijing} {\em implementation} of the TCM overpredicts low-energy scattered partons it also underpredicts corresponding jet angular correlations compared to data~\cite{anomalous}.  The \textsc{hijing} fragmentation model for low-energy partons appears to proceed by partons (gluons) $\rightarrow$ single hadrons, not correlated charge-neutral hadron pairs as observed for instance in \pp\ collisions~\cite{porter2,porter3}. In contrast, measured \pp\ dijet production is fully consistent with TCM centrality trends for jet-related structure in more-peripheral \auau\ collisions~\cite{anomalous}.

\subsection{In-vacuum jet phenomenology}

A primary goal of TCM analysis is direct comparison between measured spectrum and correlation hard components and pQCD jet systematics. The main source of isolated (in-vacuum) jet properties is \ep\ and \ee\ data from the HERA and LEP, both dijet fragment multiplicities~\cite{jetmult} and fragment momentum distributions or {\em fragmentation functions} (FFs)~\cite{tasso,opal}. FF systematics are conveniently plotted on the space $(y,y_{max})$, where $y \equiv \ln[(E+p)/m_\pi]$ and $y_{max} \equiv \ln(Q/m_\pi)$ with dijet energy $Q = 2E_{jet}$.  Such FF ensembles are conditional distributions $D(y:y_{max})$ on hadron fragment rapidity $y$ given some value of parton rapidity $y_{max}$. Conditional hard components isolated from TA correlations as in the present study should be directly comparable.

\begin{figure}[h]
  \includegraphics[width=1.65in,height=1.65in]{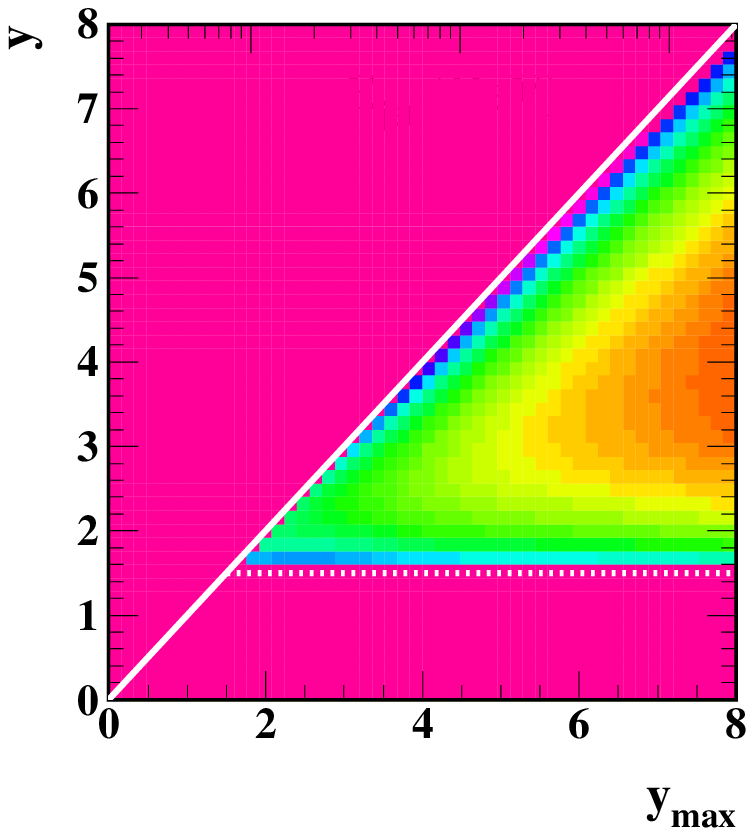}
  \includegraphics[width=1.65in,height=1.65in]{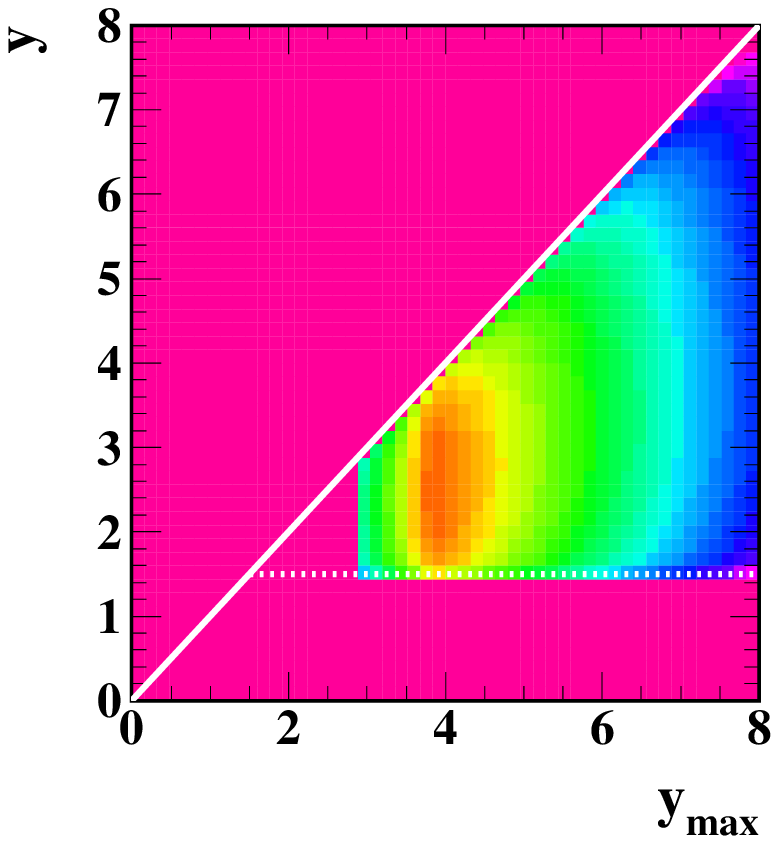}
\caption{\label{qcd}
Left: Parametrization of the fragmentation-function (FF) ensemble for unidentified hadrons from light (uds) quarks or gluons over the parton energy range 2-200 GeV ($y_{max} \in$ [3.3,8]), from Ref.~\cite{eeprd}.
Right: The same FF ensemble folded with a perturbative QCD parton spectrum on $y_{max}$ as in Ref.~\cite{fragevo}. The z axis is logarithmic in both cases.
 }  
 \end{figure}

Figure~\ref{qcd} (left) shows a parametrization of measured FFs~\cite{eeprd} from the lowest observed jet energies ($\approx 2$ GeV or $y_{max} \approx 3.3$ from Ref.~\cite{cleo}) to the highest ($\approx 100$ GeV or $y_{max} \approx 8$ from LEP). The lower bound on $y$ is determined by \ppbar\ FF data~\cite{fragevo}. The parametrization has the simple form $D(y:y_{max}) = 2n_{ch}(y_{max}) \beta(u;p,q)$, where $u \approx y/y_{max}$ is a scaled rapidity, $\beta(u;p,q)$ is the unit-normal beta distribution, $n_{ch}(y_{max})$ is determined by parton energy conservation within a jet, and parameters $(p,q)$ are nearly constant over a large parton energy interval. That simple parametrization describes all quark or gluon $\rightarrow$ unidentified hadron FFs over a large energy interval to the uncertainty limits of the FF data.

Figure~\ref{qcd} (right) shows the \ee\ FF ensemble in the left panel multiplied by a pQCD parton (power-law) spectrum on $y_{max}$~\cite{fragevo}. The most-probable jet (minijet) energy is $E_{jet} \approx 3$ GeV ($y_{max} \approx$ 3.75) consistent with measured jet systematics~\cite{ua1,anomalous,fragevo}.  That distribution is the argument of a pQCD folding integral (factorization, Bayes' theorem). Projection onto the vertical axis (integration over the parton spectrum on $y_{max}$) produces a {\em fragment distribution}---a pQCD prediction for the measured TCM spectrum hard component $H(y_t)$~\cite{fragevo}. The hard components of 2D TA correlations may be compared quantitatively with the FF data in Fig.~\ref{qcd}. Such comparisons should establish the kinematic limits for jet manifestations in high-energy nuclear collisions.

\subsection{Symmetrized two-particle correlations}

Correlation structures on $(\eta_\Delta,\phi_\Delta)$ and $y_t \times y_t$ from 200 GeV NSD \pp\ collisions~\cite{porter2,porter3} are consistent with extrapolation of the centrality systematics of angular correlations from \auau\ collisions~\cite{anomalous}. Both angular correlations and correlations on transverse rapidity are described accurately  by the TCM (aside from small-amplitude structures representing Bose-Einstein correlations  and gamma conversion to \ee\ pairs).

The soft component of 2D angular correlations is represented by a narrow 1D Gaussian on $\eta_\Delta$. Soft-component pairs are exclusively the unlike-sign (US) charge combination consistent with projectile-proton dissociation (mainly to gluons near mid-rapidity), and the r.m.s.\ peak width on $\eta_\Delta$ is $\sigma_{soft} \approx 0.5$~\cite{porter2}.
The hard-component 2D peak  on $y_t \times y_t$ corresponds approximately (when projected onto 1D $y_t$) to the 1D SP spectrum hard component $H_0(y_t)$ of Eq.~(\ref{ppspec}). The hard component of 2D angular correlations (two features) is consistent with expectations for intra-jet correlations (SS 2D peak, ``jet cone'') and inter-jet correlations (AS 1D peak on azimuth, back-to-back jets). The volume of the SS 2D peak corresponds quantitatively to the hard component of the total hadron yield inferred from $y_t$ spectrum data and to pQCD calculations~\cite{jetspec}. Measured \pp\ angular correlations~\cite{porter2} show only semiquantitative agreement with the \textsc{pythia} Monte Carlo~\cite{pythia}. 

 \subsection{Conditional trigger-associated correlations}

Single-particle spectrum hard components (jet fragment distributions) are 1D projections of more-complex extended objects (dijets). We seek to isolate 2D trigger-associated hard components that reveal further details of jet structure, especially the analog between TA conditional distributions on space $(y_{ta},y_{tt})$ inferred from \pp\ collisions and parton-fragment conditional distributions on space $(y,y_{max})$ inferred from \ee\ collisions~\cite{eeprd,fragevo}. 

The TA pair distribution for given $n_{ch}$ class is denoted by $F(y_{ta},y_{tt},n_{ch})$. To extract a 2D hard component from TA data we require a 2D soft-component model. Equation~(\ref{ppspec}) is a TCM for SP  \yt\ spectra from which the 1D hard component $H(y_t)$ was isolated. Deriving a TCM for the 2D TA distribution is an exercise in compound probabilities and random sampling from multiple parent distributions. Trigger and associated particles are approximately random samples from fixed spectrum components $S_0(y_t)$ and $H_0(y_t)$ in Eq.~(\ref{ppspec}). 

We assume that the soft component of the TA TCM is factorizable (uncorrelated samples). We further assume, only for purposes of illustration in this study, that the 2D hard-component {\em model} is also factorizable. We expect that the TA {\em data} hard component derived by subtracting the soft-component model is not factorizable in general. 

We factorize the 2D TA distribution for given $n_{ch}$ class according to Bayes' theorem as
\bea \label{ta2comp}
F(y_{ta},y_{tt}) &=& T(y_{tt}) A(y_{ta}:y_{tt}),
\eea
where $T(y_{tt})$ is a {\em trigger spectrum} and $A(y_{ta}:y_{tt})$ is an ensemble of conditional {\em associated-particle spectra} given a trigger at $y_{tt}$.  We first obtain a TCM for the 1D trigger spectrum and then develop the 2D TA TCM as a Cartesian product.

\section{One-dimensional Trigger spectra} \label{trig}

Trigger particles arise from (sample) soft or hard events, and for hard events arise from either the soft or hard spectrum component. For a given trigger particle the remainder of an event with multiplicity $n_{ch}$ is constrained to $n_{ch} - 1$ associated particles. We now combine information from Refs.~\cite{ppprd,fragevo} on dijet frequency and the two-component 1D \yt\ spectrum model to derive a 1D TCM for trigger spectrum $T(y_{tt},n_{ch})$. The total multiplicity in the angular acceptance is $n_{ch} = \Delta \eta\,  dn_{ch}/d\eta$.

For a sequence of {\em independent} trials (collision events), each including $n_{ch}$ samples from a fixed parent spectrum $F_x(y_t)$ ($x$ denotes soft $s$ or hard $h$ event type), we sort events according to the maximum sample value $y_{tt}$ in each event. 
Parent distributions are denoted by  $F_s(y_{tt}) = S_0(y_{tt})$ for soft events and $F_h(y_{tt},n_{ch}) = p_s' (n_{ch})S_0(y_{tt}) + p_h'(n_{ch}) H_0(y_{tt})$ for hard events, where $p_x' = n_x' / n_{ch}$.
%
We then define the trigger spectrum
\bea \label{trigspec}
T(y_{tt},n_{ch}) &\equiv& \frac{1}{N_{evt}(n_{ch})} \frac{dn_{trig}}{y_{tt}dy_{tt}} \\ \nonumber 
&& \hspace{-.3in} =  P_s(n_{ch})T_{s0}(n_{ch})G_s(y_{tt},n_{ch})\, n_{ch} F_s(y_{tt}) \\ \nonumber 
&& \hspace{-.27in} +      P_h(n_{ch})T_{h0}(n_{ch})G_h(y_{tt},n_{ch})\, n_{ch} F_h(y_{tt},n_{ch}) \\ \nonumber
&& \hspace{-.3in} = P_s(n_{ch})T_s(y_{tt},n_{ch}) + P_h(n_{ch})T_h(y_{tt},n_{ch}). 
\eea
In each term factor $F_x(y_{tt},n_{ch})$ is the probability of a $y_{tt}$ sample from the parent spectrum, and factor $G_x(y_{tt},n_{ch})$ is the probability of a void (no samples) above the given $y_{tt}$. The sum gives the overall density of trigger particles at $y_{tt}$ from either event type. 

The {\em void probability} $G_x(y_{tt},n_{ch})$ is defined as follows: For events with trigger at $y_{tt}$ no sample should appear with $y_t > y_{tt}$ (void interval). The event-wise spectrum integral {\em above} $y_{tt}$ within acceptance $\Delta \eta$ is
\bea \label{nsum}
n_{x\Sigma}(y_{tt},n_{ch}) 
 &=& \int _{y_{tt}}^\infty dy_t y_t n_{ch} F_x(y_{t})
\eea
separately for unit-normal spectra $F_x(y_t)$ from soft or hard events. The void probability for event type $x$ is $G_x(y_{tt},n_{ch}) = \exp[- \kappa n_{x\Sigma}(y_{tt},n_{ch})]$, where $O(1)$ factor $\kappa$ may account for non-Poisson (e.g., jet) correlations. $O(1)$ coefficients $T_{x0}(n_{ch})$ are defined such that the product $T_x \equiv T_{x0}(n_{ch}) G_x\, n_{ch} F_x$ is unit normal
\bea \label{g0yt}
\int_0^\infty dy_{tt} y_{tt} T_x(y_{tt},n_{ch}) &=& 1
\eea
(the number of trigger particles from any event is 1). Small deviations from $T_{x0} = 1$ may arise from incomplete $y_t$ acceptance and particle detection inefficiencies. 

 \begin{figure}[h]
  \includegraphics[width=3.3in]{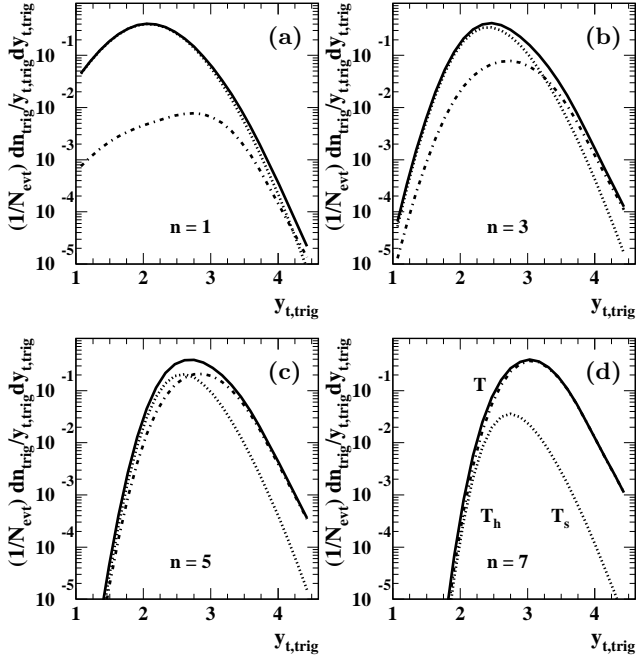}
 \put(-20,105) {\bf (d)}
 \put(-20,232) {\bf (b)}
 \put(-140,105) {\bf (c)}
 \put(-140,232) {\bf (a)}
\caption{\label{trigspec2}
Two-component models (TCM) of trigger spectra for four multiplicity classes of 200 GeV \pp\ collisions generated by Eq.~(\ref{trigspec}). Trigger spectra for soft events (no jets, dotted), hard events (at least one jet, dash-dotted) and total (solid).
 }  
 \end{figure}

Figure~\ref{trigspec2} shows trigger spectra $T(y_{tt},n_{ch})$ obtained from Eq.~(\ref{trigspec}) (solid) and the soft and hard components $P_s T_{s}$ (dotted) and $P_h T_{h}$ (dash-dotted) vs trigger rapidity  for four multiplicity classes.  TCM model parameters for $S_0(y_{tt})$ and $H_0(y_{tt})$ are those from the spectrum analysis in Ref.~\cite{ppprd} with no changes (see App.~\ref{tcmmodelfunc}).  The spectrum mode shifts to larger $y_{tt}$ values with increasing $n_{ch}$.

 \begin{figure}[h]
  \includegraphics[width=3.3in]{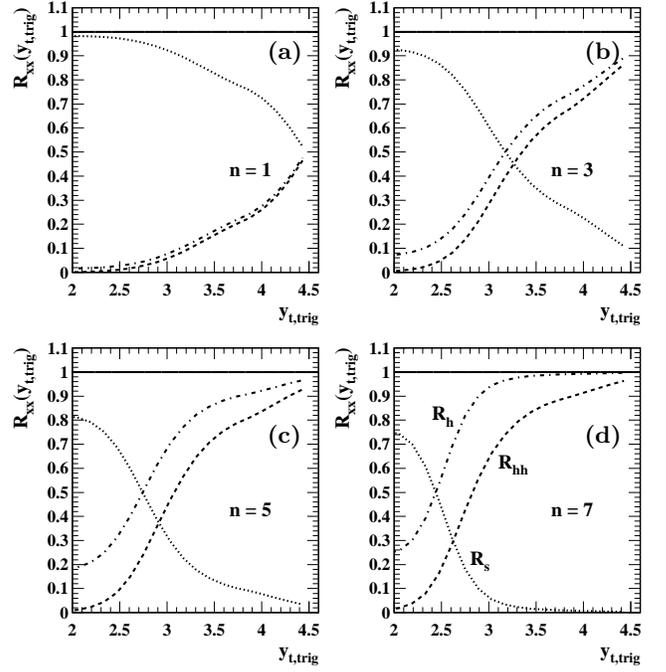}
 \put(-22,85) {\bf (d)}
 \put(-22,230) {\bf (b)}
 \put(-142,85) {\bf (c)}
 \put(-142,230) {\bf (a)}
\caption{\label{trigprobs}
TCM trigger spectrum ratios for four multiplicity classes of 200 GeV \pp\ collisions. Ratios for soft events (dotted curves), hard events (dash-dotted curves) and hard components of hard events (dashed curves). Soft/hard event-type crossings occur at 6, 2, 1 and 0.75 GeV/c respectively for the four classes.
 }  
 \end{figure}

Figure~\ref{trigprobs} shows {\em trigger fractions} $R_{s}$, $R_{h}$ and $R_{hh}$ vs trigger rapidity, where $ R_{x} = P_x T_{x} / T$ is the ratio of the corresponding $T_{x}$ term in Eq.~(\ref{trigspec}) to the full trigger spectrum $T(y_{tt},n_{ch})$. The bold dash-dotted curve is $R_h$ and the bold dotted curve is $R_s$. For higher-multiplicity events and above $p_t \approx 1$ GeV/c ($y_t \approx 2.7$) trigger particles represented by $R_{hh}$ (hard components of hard events) arise mainly from the 1D \yt\ spectrum hard component associated with jets. Those hadrons are the only proper parton proxies. Although the {\em fraction} of soft triggers for larger $n_{ch}$ still dominates at lower \yt\ the absolute number of soft  triggers is negligible.  Ratios $R_x$ and the corresponding spectra $T_x$ from Eq.~(\ref{trigspec}) play a central role in the definition of the TCM for 2D TA correlations.

 \section{Trigger-associated 2D model} \label{tata}

We now combine trigger spectrum elements from Eq.~(\ref{trigspec}) with conditional distributions based on the TCM for 1D \yt\ spectra and certain {\em marginal constraints} (described below and in App.~\ref{marg}) to define the 2D TCM $F(y_{ta},y_{tt},n_{ch})$ for TA correlations. 

\subsection{TCM TA distribution $\bf F(y_{ta},y_{tt})$}

We assume that for a given trigger the associated particles are sampled from parent distributions similar to those for the 1D SP spectrum TCM subject to additional marginal constraints described below. For multiplicity class $n_{ch}$ the total number of triggers is $N_{evt}(n_{ch})$, the associated-particle number per event is $n_{ch} - 1$, and the total trigger-associated pair number for the given $n_{ch}$ class (excluding self pairs) is $N_{evt}(n_{ch})(n_{ch}-1)$. 

We assume that the expressions for soft and hard event types are linearly independent, and event types occur with probabilities  $P_x(n_{ch})$ as defined previously. We then have $F(y_{ta},y_{tt}) = P_s F_s(y_{ta},y_{tt}) + P_h F_h(y_{ta},y_{tt})$. According to Bayes' theorem the probability of a TA pair $F_x(y_{ta},y_{tt})$ can be written as the product of trigger probability $T_x(y_{tt},n_{ch})$ and $A_x(y_{ta}:y_{tt},n_{ch})$, the conditional probability of an associated particle at $y_{ta}$ given a trigger at $y_{tt}$.
$A_x(y_{ta}:y_{tt})$ is based on the $F_x(y_t)$ from the 1D SP TCM but is set to zero above $y_{tt}$ (void). 
Combining the conditional probabilities with the corresponding  trigger-spectrum components we obtain a 2D TCM for unit-normal $F(y_{ta},y_{tt},n_{ch})$
\bea \label{tadist}
F(y_{ta},y_{tt}, n_{ch})
&=& \frac{1}{N_{evt}(n_{ch})(n_{ch}-1)}\frac{d^2n_{ch}}{y_{tt}dy_{tt}y_{ta}dy_{ta}}~~ \\ \nonumber
&=&    P_s (n_{ch}) T_s(y_{tt},n_{ch})A_s(y_{ta}:y_{tt},n_{ch}) \\ \nonumber
&+& P_h (n_{ch}) T_h(y_{tt},n_{ch}) A_h(y_{ta}:y_{tt},n_{ch}),
\eea
where $A_s(y_{ta}:y_{tt},n_{ch}) = S_0''(y_{ta}:y_{tt},n_{ch})$ for soft and $A_h(y_{ta}:y_{tt},n_{ch}) = p_s'(n_{ch}) S_0'(y_{ta}:y_{tt},n_{ch}) + p_h'(n_{ch}) H_0'(y_{ta}:y_{tt},n_{ch})$ for hard event types. Primes indicate that the conditional probabilities may deviate from the corresponding 1D SP spectrum models because of imposed marginal constraints. Modifications to the SP spectrum components are represented by $O(1)$ weight functions $D_x(y_{tt})$ and $E_x(y_{ta})$ as described in App.~\ref{marg}. Note that for associated samples $p_x' = n_x'/(n_{ch}-1)$. 

 \begin{figure}[h]
  \includegraphics[width=3.3in]{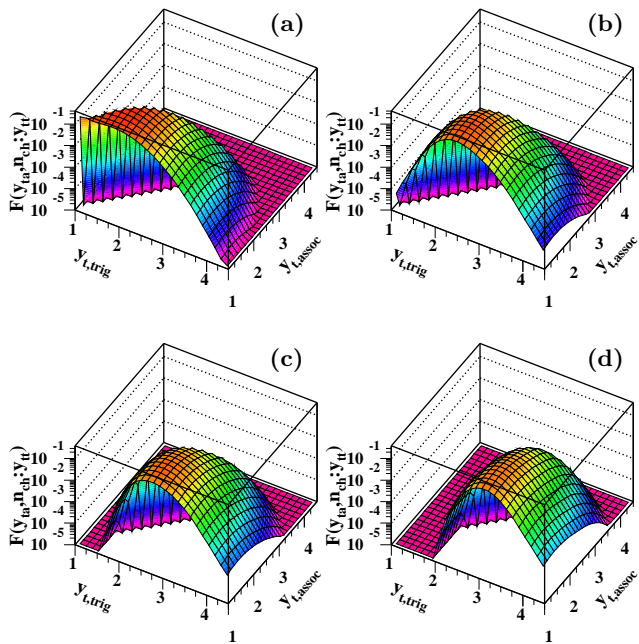}
 \put(-20,105) {\bf (d)}
 \put(-20,232) {\bf (b)}
 \put(-140,105) {\bf (c)}
 \put(-140,232) {\bf (a)}
\caption{\label{spec1a}
Two-component model for trigger-associated correlations from four multiplicity classes of 200 GeV \pp\ collisions generated by Eq.~(\ref{tadist}).
 }  
 \end{figure}

Figure~\ref{spec1a} shows the predicted TCM $F(y_{ta},y_{tt},n_{ch})$ simulating measured 2D TA correlations for four \pp\ multiplicity classes. The rapidity interval  $y_t \in [1,4.5]$ is divided into 25 histogram bins consistent with previous $y_t \times y_t$ data analysis~\cite{porter2,porter3}. The hard components of measured 2D TA distributions should relate to the jet systematics of Fig.~\ref{qcd}. The soft components of the TCM defined by Eq.~(\ref{tadist}) may be used to isolate TA data hard components for such comparisons. We now consider details of the conditional distributions $A_x(y_{ta}:y_{tt},n_{ch})$.

\subsection{TCM marginal constraints}

The 2D TA distribution $F(y_{ta},y_{tt},n_{ch})$ must integrate over $y_{ta}$ to the unit-normal trigger spectrum $T(y_{tt},n_{ch})$ and over $y_{tt}$ to the unit-normal SP spectrum $F(y_{ta},n_{ch})$ (minus the trigger spectrum if self pairs are excluded). Those marginal constraints modulate the conditional distributions  $A_x(y_{ta}:y_{tt})$ in the 2D TA TCM and must be met separately for soft and hard events. We seek the most general 2D model functions that satisfy the constraints. 

For event-type $x$ a sum over all events (triggers) should return the 1D parent spectrum $F_x(y_t)$, but we exclude self pairs (triggers) from the 2D TA distribution. Since the trigger self pairs appear along the main diagonal they are described on $y_{ta}$ by the same function $T_x(y_{ta},n_{ch})$. The complement is the unit-normal marginal spectrum of associated particles $A_x(y_{ta})$ defined for given $n_{ch}$ by 
\bea
(n_{ch}-1)A_x(y_{ta}) =n_{ch} F_x(y_{ta}) - T_x(y_{ta}).
\eea
The following must be true for independent sampling from SP parent distribution $F_x(y_t)$ for given $n_{ch}$
\bea \label{f0yt}
A_x(y_{ta}) &=& \int_{y_{ta}}^\infty dy_{tt} y_{tt} T_x(y_{tt}) A_x(y_{ta}:y_{tt}).
\eea

Each associated-particle conditional probability should be unit normal on $y_{ta}$ for any $y_{tt}$
\bea \label{f0yt2}
\int_0^{y_{tt}} dy_{ta} y_{ta} A_x(y_{ta}:y_{tt}) &=& 1,
\eea
so that integration of Eq.~(\ref{f0yt}) over $y_{ta}$ returns Eq.~(\ref{g0yt}).
Equations~(\ref{f0yt}) and (\ref{f0yt2}) define the required marginal constraints on the 2D conditional probability distributions. 

Conditional distributions $A_x(y_{ta}:y_{tt})$ can be obtained by an iterative process. We refer to histogram $A_x(y_{ta}:y_{tt})$ as having rows indexed by $y_{ta}$ and columns indexed by $y_{tt}$. To form the initial approximation to $A_x(y_{ta}:y_{tt})$ 1D  distribution $F_x(y_{ta})$ from the SP TCM is set to zero above $y_{tt}$ for each $y_{tt}$. Equation~(\ref{f0yt2}) is imposed for each $y_{tt}$ to renormalize the corresponding column. Equation~(\ref{f0yt}) is then imposed for each $y_{ta}$ to renormalize the corresponding row, which may change the column normalizations. The resulting alteration of SP model functions $F_x$ is represented by weight functions $D_x$ and $E_x$. Iteration converges rapidly and a single pass provides sufficient accuracy within the kinematic region on $(y_{ta},y_{tt})$ relevant to the TA hard component. Further details are presented in App.~\ref{marg}.

 \begin{figure}[h]
  \includegraphics[width=1.65in]{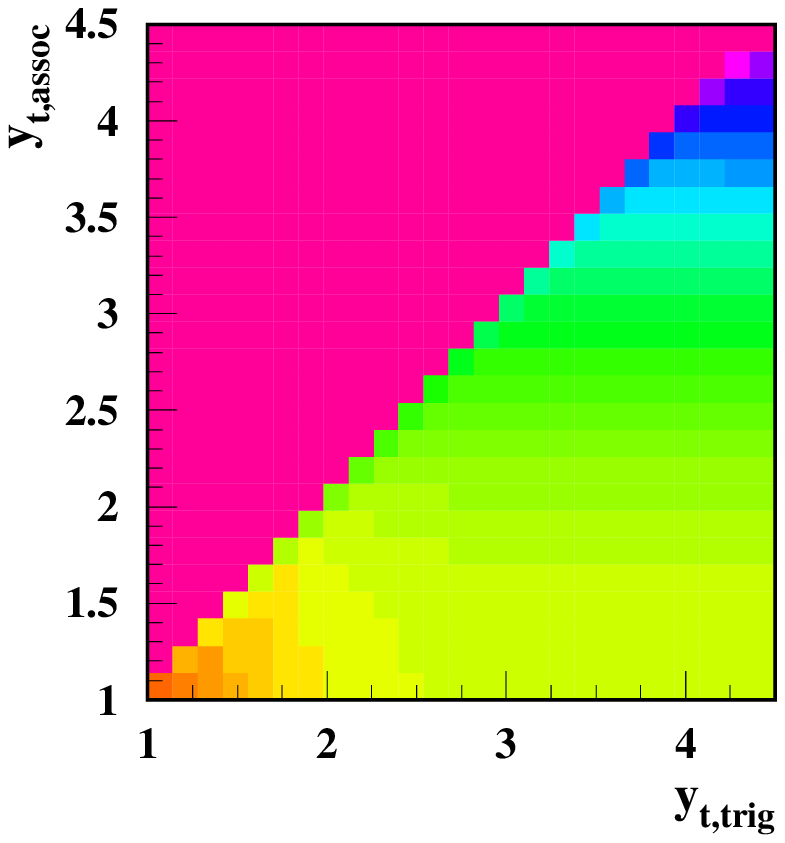}
  \includegraphics[width=1.65in]{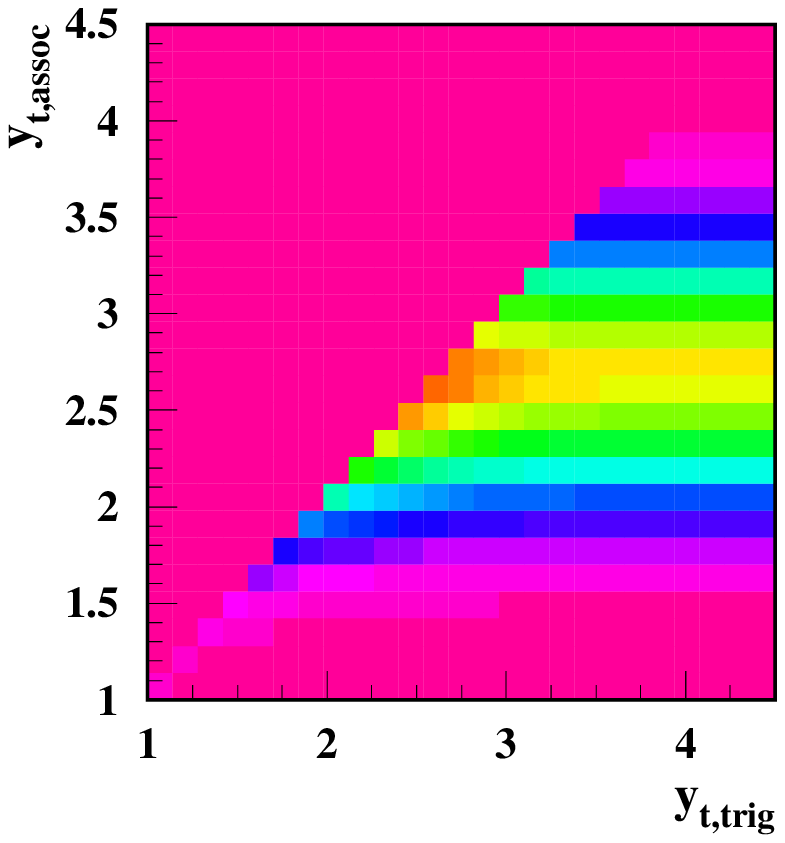}
\caption{\label{spec3aa}
Left: Soft-event conditional distribution $A_s(y_{ta}:y_{tt}) = S_0''(y_{ta}:y_{tt})$ showing the effects of weight factor $D_s(y_{tt})$ at smaller $y_{tt}$ as described in App.~\ref{marg}. The z axis is logarithmic.
Right:  Hard-event hard component conditional distribution $H_0'(y_{ta}:y_{tt},n_{ch})$ showing the effects of weight function $D_h(y_{tt})$ at smaller $y_{tt}$. The z axis is linear.
 }  
 \end{figure}

Figure~\ref{spec3aa} illustrates the effects of marginal constraints on $S_0''(y_{ta}:y_{tt})$  (left) and $H_0'(y_{ta}:y_{tt})$ (right). The initial $A_x(y_{ta}:y_{tt})$ forms follow the SP parent spectra $F_x(y_{t})$ on $y_{ta}$ and are uniform on $y_{tt}$ subject to the condition $y_{ta} < y_{tt}$. The constraint of Eq.~(\ref{f0yt2}) represented by $D_x(y_{tt})$ causes elevated amplitudes at smaller $y_{tt}$ due to the reduced acceptance on $y_{ta}$. The constraint  of Eq.~(\ref{f0yt}) represented by $E_x(y_{ta})$ might alter amplitudes at larger $y_{ta}$ relative to the SP parents. Any renormalization on $y_{ta}$ could then disturb the normalization on $y_{tt}$ requiring multiple iterations for convergence, but we observe that such effects are negligible. Systematic  uncertainties for the renormalizations are discussed in Sec.~\ref{syserr}.

\subsection{TCM conditional distribution $\bf A(y_{ta}:y_{tt})$}

The full TA distribution can be factorized according to Bayes' theorem to define a complementary conditional distribution. 
We divide $F(y_{ta},y_{tt},n_{nch})$ from Eq.~(\ref{tadist}) by trigger spectrum $ T(y_{tt},n_{ch})$ from Eq.~(\ref{trigspec}) to define a TCM for measured $A(y_{ta}:y_{tt},n_{ch})$. The result is an ensemble of unit-normal conditional spectra
\bea \label{ftrat}
A(y_{ta}:y_{tt},n_{ch}) &=&  \frac{1}{(n_{ch}-1)} \frac{dn_{ch}(y_{ta}:y_{tt},n_{ch})}{y_{ta}dy_{ta}} \\ \nonumber
&=&   R_{s}(y_{tt},n_{ch}) A_s(y_{ta}:y_{tt},n_{ch}) \\ \nonumber
&+&    R_{h}(y_{tt},n_{ch})  A_h(y_{ta}:y_{tt},n_{ch}).
\eea

 \begin{figure}[h]
  \includegraphics[width=3.3in]{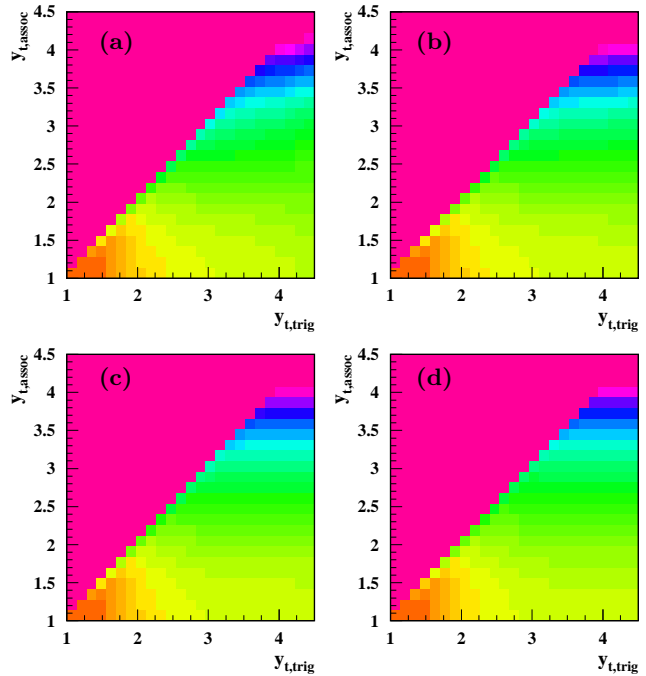}
 \put(-85,110) {\bf (d)}
 \put(-85,237) {\bf (b)}
 \put(-205,110) {\bf (c)}
 \put(-205,237) {\bf (a)}
\caption{\label{spec3a}
Two-component model for trigger-associated conditional spectra from four multiplicity classes of 200 GeV \pp\ collisions generated by Eq.~(\ref{ftrat}). The jet-related hard-component contribution at upper right increases with $n_{ch}$. The z axis is logarithmic.
 }  
 \end{figure}

Figure~\ref{spec3a} shows TA histograms from Fig.~\ref{spec1a} divided by trigger spectra from Eq.~(\ref{trigspec}) to obtain TCM TA histograms $A(y_{ta}:y_{tt},n_{ch})$. As noted, the increase in amplitude with decreasing $y_{tt}$ for smaller $y_{tt}$ occurs because each conditional distribution must be unit normal on $y_{ta}$ but extends over a decreasing interval $y_{ta} \in[1,y_{tt}]$.  


\section{Isolating the TA hard component} \label{isolate}

Conditional ratio $A(y_{ta}:y_{tt})$ obtained directly from measured TA histograms $F$ divided by measured trigger spectra $T$ is model independent and should provide the least-biased route to sought-after jet-related structure.  Isolation of the hard component requires  subtraction of a soft-component model (part of the TA TCM). 

\subsection{Hard components of $\bf A(y_{ta}:y_{tt},n_{ch})$}

The 2D TA histograms in the previous section can be constructed from measured particle data. We wish to isolate a measured TA hard component  $H_h(y_{ta}:y_{tt},n_{ch})$ by subtracting a 2D soft-component  reference according to the method described in Ref.~\cite{ppprd}. Solving Eq.~(\ref{ftrat}) for the hard component of {\em hard events} we obtain
\bea \label{hardcomp}
H_h'(y_{ta}:y_{tt})/(n_{ch}-1) &=& [A - R_s S''_0]/R_h -  p'_s S'_0~~
\eea
where for real data $A_s(y_{ta}:y_{tt}) \equiv S_0''(y_{ta}:y_{tt})$ and $ A_h(y_{ta}:y_{tt}) \equiv p'_s S_0' + H_h'(y_{ta}:y_{tt})/(n_{ch}-1)$. The primes on the $F_x$ recall the effects of marginal constraints on the TCM (App.~\ref{marg}). Subscript $h$ denotes a component from hard events as in Eq.~(\ref{spevtype}).  For this model exercise we expect the factorization result $H_h'(y_{ta}:y_{tt},n_{ch})  \rightarrow n_{h}' H'_0(y_{ta}:y_{tt})$.

 \begin{figure}[h]
  \includegraphics[width=3.3in]{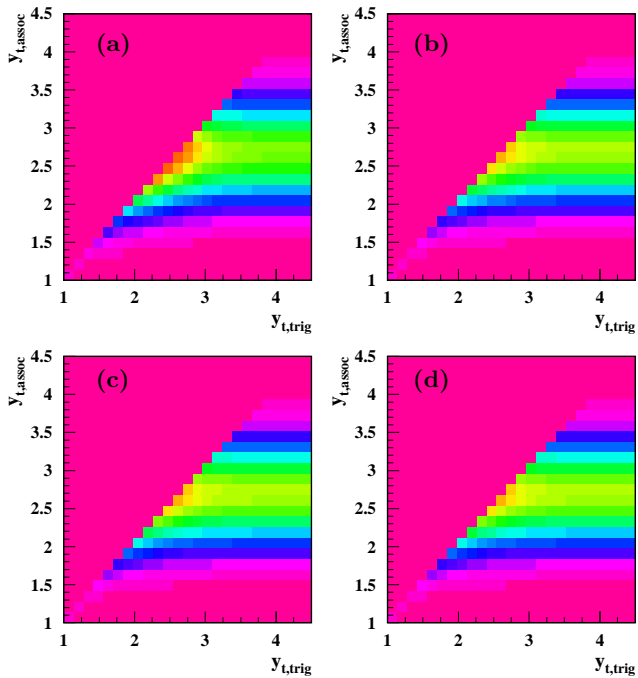}
 \put(-85,110) {\bf (d)}
 \put(-85,237) {\bf (b)}
 \put(-205,110) {\bf (c)}
 \put(-205,237) {\bf (a)}
\caption{\label{hardcol1}
Hard components of trigger-associated conditional spectra from four multiplicity classes of 200 GeV \pp\ collisions in the form $H_h'(y_{ta}:y_{tt})/n_{h}'$ derived according to Eq.~(\ref{hardcomp}). The procedure returns the hard component $H'_0(y_{ta}:y_{tt})$ of the TCM approximately independent of $n_{ch}$ as expected. The z axis is linear.
 }  
 \end{figure}

Figure~\ref{hardcol1} confirms that model hard component $H_0'(y_{ta}:y_{tt})$ is returned by soft-component subtraction applied to the simulated data in Fig.~\ref{spec3a} for each $n_{ch}$ class (indicating minor variation of weight function $D_h$ with $n_{ch}$ as in Fig.~\ref{specc4}, upper right). The TA TCM assumes factorization for the hard component. For real data we expect to encounter the primary goal of 2D TA analysis: nontrivial jet-related correlation structure in $H_h(y_{ta}:y_{tt},n_{ch})$ reflecting low-energy parton fragmentation systematics.

We can compare a limiting case of the 2D TA system with the 1D SP spectrum analysis. For larger $y_{tt}$ and $n_{ch}$ we expect $R_s \ll R_h \approx 1$ (see Fig.~\ref{trigprobs}) and 
\bea
H_h'(y_{ta}:y_{tt}) \hspace{-.05in} &\approx&\hspace{-.05in} (n_{ch}-1) A(y_{ta}:y_{tt}) \hspace{-.01in} - \hspace{-.01in} n_s' S_0'(y_{ta}:y_{tt}),~~~~~
\eea
equivalent to the SP spectrum problem described in Sec.~\ref{ppspecc}, with $P_s \ll P_h \approx 1$. 
For ensemble-averaged SP spectra we observe $H_h(y_t,n_{ch}) \approx n_h'(n_{ch})  H_0(y_t)$ with $n_h'(n_{ch}) \approx \bar n_{ch,j}$ -- factorization of $H_h(y_t,n_{ch})$. For measured TA data, jet-related $H_h(y_{ta}:y_{tt},n_{ch})$ ($H_h'$ corrected for constraint distortions) may not factorize. The 2D structure should correspond to Fig.~\ref{qcd} and may reveal further details of minimum-bias jets in p-p collisions.

\subsection{Hard components of $\bf F(y_{ta},y_{tt},n_{ch})$}

 \begin{figure}[h]
  \includegraphics[width=3.3in]{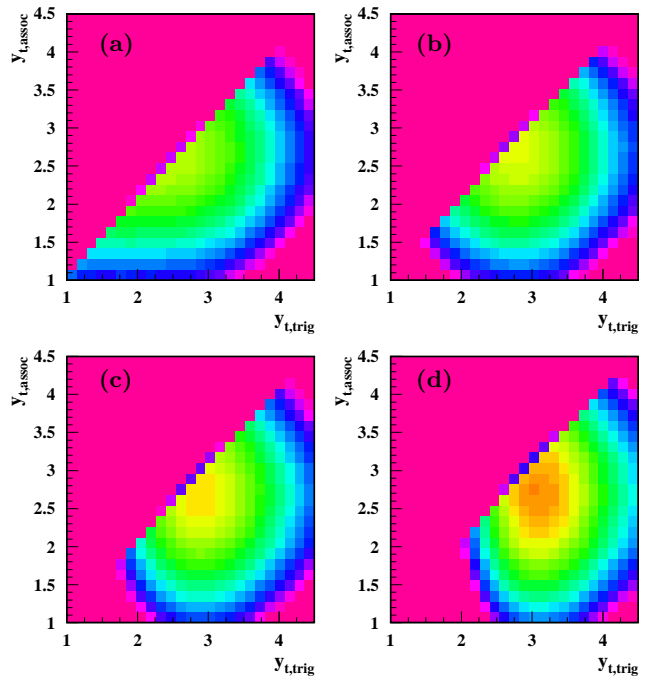}
 \put(-85,110) {\bf (d)}
 \put(-85,237) {\bf (b)}
 \put(-205,110) {\bf (c)}
 \put(-205,237) {\bf (a)}
\caption{\label{hardcol2a}
Hard components of minimum-bias trigger-associated correlations from Fig.~\ref{spec1a} for four multiplicity classes of 200 GeV \pp\ collisions. These distributions were reconstructed by combining hard components $H'(y_{ta}:y_{tt})$ from Fig.~\ref{hardcol1} with model trigger-spectrum hard components $T_h(y_{tt})$ from Eq.~(\ref{trigspec}). The z axis is logarithmic.
 }  
 \end{figure}

Figure~\ref{hardcol2a} shows the product $T_h(y_{tt},n_{ch}) H_h'(y_{ta}:y_{tt},n_{ch})/n_h'$ representing the form of the (factorized) hard component of TA distributions $F(y_{ta},y_{tt},n_{ch})$ in Fig.~\ref{spec1a}. The distribution mode on $y_{tt}$ moves to larger $y_{tt}$ with increasing $n_{ch}$.  For large multiplicities the mode approaches 3 GeV ($y_{tt} \approx 3.75$), the observed most-probable parton energy. The mode for lower multiplicities is closer to $y_t = 2.7$ observed for minimum-bias symmetrized $y_t \times y_t$ data from 200 GeV \pp\ collisions~\cite{porter2,porter3}. In this simulation exercise the distributions extend below $y_{ta} = 2$. We expect systematic uncertainties in inferred hard components for data to be large below that point.

 \section{Systematic uncertainties} \label{syserr}

The object of proposed trigger-associated correlation analysis is extraction of a jet-related hard component that can be compared with measured in-vacuum jet properties and a predicted pQCD parton spectrum. The accuracy of the inferred TA hard component depends on possible limitations of the analyzed data and the validity of both the overall TA two-component description and of the defined soft-component model function.

\subsection{Validity of the trigger-associated TCM}

The predicted TA TCM is based on the measured TCM for single-particle spectra ($S_0$ and $H_0$ defined in App.~\ref{tcmmodelfunc}) which has been verified in comparisons with spectrum data at the few-percent level~\cite{ppprd}. The structure of the TA model is $F = P_s T_s S''_0 + P_h T_h (p_s' S'_0 + p_h' H'_0)$. The $n_{ch}$ trends from SP spectra determine soft and hard particle probabilities $p_x'$ with few-percent accuracy. Modifications to the SP model components denoted by primes are discussed in App.~\ref{marg} and should not contribute significantly to uncertainties in the inferred TA hard component.
The probabilities $P_x$ of soft and hard events depend on the inference of jet frequency $f$ from spectrum data, which in turn depends on interpretation of the SP spectrum hard component as a jet fragment distribution. That interpretation has been confirmed quantitatively in Ref.~\cite{fragevo}.

For a given event type the corresponding TA distribution $F_x(y_{ta},y_{tt})$ is factorized according to Bayes' theorem to a product of trigger spectrum and associated-particle conditional spectra. Trigger spectra are defined by product $T_x = G_x n_{ch} F_x$ reflecting the probability that a given sample (the trigger) occurs at $y_{tt}$ and that no other sample occurs above that point (the void). The predicted $T(y_{tt},n_{ch})$ can be compared directly with measured trigger spectra as a critical test of the compound-probability analysis that is the basis for the TA TCM.

\subsection{TA TCM model elements} 

The TA hard component must be isolated from TA correlation data by subtracting a 2D soft-component model with contributions from both soft and hard event types. Elements of the 2D soft component are expressed as $T_s S_0''$ (soft events) and $T_h p_s' S_0'$ (hard events). We assume that each element is factorizable (no correlations between trigger and associated particles, both from soft processes).

The TA hard-component model is introduced in this study only as a place holder to verify the subtraction procedure -- that the hard component included in the TCM emerges from subtraction with only expected small distortions due to marginal constraints. The TA hard-component model is factorized and thus cannot represent the complex fragmentation conditional systematics that appear in Fig.~\ref{qcd} and are expected for TA real data. Jet-related correlations are the main objective of TA analysis.

\subsection{Common-mode uncertainty reduction}

By inferring the TA hard component from ratio $A = F / T$ certain common-mode reductions in systematic bias may be achieved. Conditional associated-particle spectra have the form $A = R_s S_0'' + R_h (p_s' S_0' + H'/n_{ch})$, where $F$, $T$ and $A$ are obtained directly from TA data. We expect the hard-component structure to be slowly varying with $n_{ch}$ (the jet formation process is approximately independent of \pp\ multiplicity). For larger $n_{ch}$ and for $y_{tt} > 2.5$ ($p_t > 0.8$ GeV/c) $R_s \ll R_h \approx 1$. Uncertainty in the inferred hard component $H'$ then arises only from $p_s'$, which is accurately known, from the forms of $S_0''$ and $S_0'$ which are determined from measured SP spectrum structure and from marginal constraints reviewed in App.~\ref{marg}. 

\subsection{Data hard component inferred by subtraction}

Based on results for SP spectrum data as in Fig.~\ref{ppspectra} we expect the jet-related hard component to be resolvable (systematic uncertainties $< 10$\%) only for $y_t > 2$~\cite{ppprd}. Uncertainties contributed by the soft-component model fall sharply above that point since the model is rapidly decreasing while the hard component is increasing toward a maximum near $y_t = 2.7$ ($p_t = 1$ GeV/c). The relevant kinematic region for the 2D TA hard component is then defined by both $y_{ta}$ and $y_{tt} > 2$ ($p_t > 0.5$ GeV/c). We then summarize uncertainties relevant to that region.

The soft-component model is assumed to be factorized (no correlations). Soft-component correlations are observed in \pp\ data, but only below $y_t = 2$~\cite{porter2}, and so should not affect isolation of the TA hard component within the relevant kinematic region. 

The soft-component model is distorted by marginal constraints (mainly the projection onto $y_{tt}$ represented by $D_x$). However, the effect is significant only below $y_{tt} = 2.5$ (Fig.~\ref{specc4}, upper panels) and can be corrected to a few percent. Similarly, distortions of the soft component on $y_{ta}$ should be less than 20\% (Fig.~\ref{specc4}, lower panels) for any $y_{ta}$. We also note from the same figure that any corrections to the inferred hard component [$H_h'(y_{ta}:y_{tt}) \rightarrow H_h(y_{ta}:Y_{tt})$] in response to marginal-constraint distortions would be $< 15$\%, with few-percent accuracy of corrected data.

The $n_{ch}$ dependence of corrected $H_h(y_{ta}:y_{tt},n_{ch})$ could provide an important check on the overall method. As  noted, the hard component may represent universal parton fragmentation to low-energy jets which should be approximately independent of the \pp\ underlying event.

 \section{Discussion}  \label{disc}

In this study we have derived a two-component model of trigger-associated correlations for \pp\ collisions. Subtraction of the model soft components from TA data should reveal a 2D hard component that may establish the kinematic limits of jet production in \pp\ (and possibly \aa) collisions. We discuss the relation between \pp\ TA correlations and other aspects of nuclear collisions.

\subsection{Relation to previous MB correlation analysis}

Trigger-associated correlation analysis is an extension of previous measurements of minimum-bias angular and $y_t \times y_t$ correlations~\cite{porter2,porter3,anomalous} to develop a more complete description of minimum-bias jets in nuclear collisions. From correlation studies of \pp\ collisions we observe that the soft component of angular correlations is restricted to $y_t < 2$ ($p_t < 0.5$ GeV/c) and appears only for unlike-sign (US) charge pairs. Similarly, the jet-related same-side correlation peak is dominated by US pairs if $y_t < 4$ ($p_t < 4$ GeV/c). Both trends are consistent with local charge conservation expected during fragmentation of low-energy partons (gluons) to hadrons at the end of any fragmentation cascade. This study establishes the basic algebraic relations. TA analysis of real data may address separately the various charge combination LS, US, CI and CD (Sec.~\ref{symmcorr}) to confirm correspondence of the TA hard component with parton fragmentation. The dependence of TA structure on azimuth relative to the trigger direction can also be studied to confirm a jet interpretation of certain MB angular correlation structure. 

\subsection{pQCD and parton fragmentation functions}

We have asserted that the TA hard component may be compared in some sense to Fig.~\ref{qcd}. In Fig.~\ref{qcd} (left) the normalization for conditional distribution $D(y:y_{max})$ at each $y_{max}$ is the corresponding dijet multiplicity $2n_{ch,j}(y_{max})$~\cite{eeprd}, the ``associated particle'' (fragment) multiplicity emerging from a pair of (trigger) partons. TA associated-particle spectra  $A_h(y_{ta}:y_{tt},n_{ch})$ from hard events include soft and hard components. The normalization is the same associated-particle multiplicity $n_{ch}-1$ for all $y_{tt}$ conditions, but the fraction of $n_{ch} - 1$ from the soft component of hard events is unrelated to the dijet. For larger $n_{ch}$ soft component $n_s$ should dominate $n_{ch}$ and the $n_{ch} - 1$ constraint on the total associated-particle spectra may have only a small effect on (bias) hard component $H_h(y_{ta}:y_{tt},n_{ch})$. Thus, $D(y:y_{max})$ and $H_h(y_{ta}:y_{tt},n_{ch})$ may be directly comparable.

In Fig.~\ref{qcd} (right) the FF ensemble $D(y:y_{max})$ is combined with a calculated parton spectrum. The corresponding result for TA correlations is shown in Fig.~\ref{hardcol2a}. It may be possible to relate the hard component $T_{hh}$ of the hadron trigger spectrum to the pQCD parton spectrum and FFs using the compound-probability methods employed in this study to predict the trigger-hadron spectrum, similar to the analysis of Ref.~\cite{fragevo}

\subsection{Conventional spectrum and correlation analysis}

Nominally jet-related spectrum structure and angular correlations have been studied extensively in \aa\ collisions at RHIC in the search for jet modifications in a dense QCD medium. Conjectured modifications include high-\pt\ jet suppression (spectra)~\cite{starraa} and low-energy jet (minijet) thermalization in an opaque medium (ZYAM analysis of azimuth correlations)~\cite{starzyam}.  Conventional analysis invokes restrictive trigger-associated \pt\ cuts based on questionable assumptions about jet structure, including the assumption that only ``high-\pt'' hadrons can be associated with jets. 

By establishing the actual kinematic boundaries for jet fragment production the proposed TA analysis may counter some assumptions that support conventional data analysis. Whereas it is commonly assumed that hadrons below $p_t = 2$ GeV/c emerge from a thermalized bulk medium the TA hard component may confirm that a substantial fraction of hadrons in that \pt\ interval are part of a significant jet-correlated contribution. Imposing restrictive \pt\ cuts in ZYAM analysis then may {\em exclude most jet fragments} from nominal jet analysis~\cite{tzyam}. Interpreting spectrum structure in that interval as determined by bulk medium properties (e.g., radial flow) may erroneously assign a flow interpretation to jet structure.

\subsection{p-p underlying-event studies}

In Refs.~\cite{cdfue,cmsue} measurements of $N_\perp$ (hadron yield within the azimuth {\em transverse region} or TR centered at $\pi/2$ relative to the trigger) vs trigger condition $p_{t,trig}$ and $N_\perp(p_t)$  spectra are employed to characterize the UE. The TR is expected to exclude contributions from the triggered dijet. Extrapolation of the $N_\perp$ spectrum to $p_t = 0$ is interpreted to indicate an excess yield within the TR relative to that expected for a beam-beam contribution (projectile dissociation or soft component). Substantial increases of $N_\perp$ with higher $p_{t,trig}$ values relative to the minimum-bias or non-single diffractive (NSD) multiplicity are also interpreted to reveal novel contributions to the UE, including {\em multiple parton interactions} (MPI).

Measurements of MB dijet properties~\cite{porter2,porter3} indicate that jet-related SS and AS peaks strongly overlap on azimuth, contradicting UE assumptions about exclusion of the triggered dijet from the TR~\cite{pptheory} and assumptions common to ZYAM analysis about no significant SS and AS peak overlap~\cite{tzyam}. TA analysis applied to limited intervals on azimuth relative to the trigger (``toward'' and ``away'' regions as well as the TR defined for UE analysis) may confirm a substantial triggered dijet contribution to the TR and the momentum structure of that contribution.

\subsection{QCD Monte Carlos for p-p collisions}

Monte Carlo (MC) simulations commonly employed to describe \pp\ conditions (e.g., \textsc{pythia} and \textsc{herwig}) invoke critical physical assumptions, including a fixed projectile-proton parton distribution function, a scattered-parton spectrum with assumed lower bound, a parton fragmentation model and the eikonal model for collisions of composite projectiles. One outcome of such MCs is the prediction of MPI as a substantial contribution to the UE (within the TR). Such predictions may be questioned.

The effective dijet total cross section implied by the parton spectrum lower bound assumed for a \pp\ Monte Carlo may exceed measured MB dijet production in \pp\ collisions by a factor 10 or more~\cite{liwang}. But jet-related correlations predicted by the same Monte Carlo may fall well below those actually observed in \pp\ and peripheral \aa\ collisions~\cite{anomalous}, casting doubt on the MC hadronization model. MPI contributions attributed to $N_\perp$ data from some UE analysis may actually be part of the triggered dijet that should be expected based on measured MB jet properties. Improved understanding of \pp\ collisions may result from detailed comparisons of MCs with dijet rates inferred directly from \pt\ spectra, with measured MB angular and $y_t \times y_t$ correlations and with results from TA analysis as proposed in the present study.

\section{Summary} \label{summ}

We have derived a two-component (soft+hard) model (TCM) for 2D trigger-associated (TA) correlations from 200 GeV \pp\ collisions. The model is based on a TCM for 1D single-particle (SP) \yt\ spectra. The elements of the 1D spectrum model are combined as compound probabilities to construct the 2D TA model.

TA correlations are constructed as averages of pair distributions from single events where the highest \pt\ or \yt\ (transverse rapidity) particle (trigger) in each event is combined with all other particles (associates) to form TA pairs. By subtracting the soft component of the TA TCM from TA correlation data we extract the TA hard component, which should be dominated by jet-related structure.

The projection of the TA TCM onto trigger rapidity $y_{t,trig}$ is the trigger spectrum $T({y_{tt}})$. A trigger spectrum can also be formed directly from data and compared with the TCM prediction. Projection of the TA TCM onto associated rapidity $y_{t,assoc}$ should return the SP associated spectrum $A(y_{ta})$ (SP spectrum without trigger particles, which are excluded as self pairs).

A trigger hadron from a \pp\ collision acts (with some probability) as proxy for the leading parton of a jet. Trigger-associated hadron correlations then include jet correlations that emulate parton-fragment correlations. TA conditional hard component $H_h(y_{ta}:y_{tt})$ may be compared directly with measured fragmentation functions $D(y:y_{max})$, where $y$ is the hadron fragment rapidity and $y_{max}$ is the parton rapidity. Such comparisons should establish kinematic limits on parton energy and fragment momentum for jet production in \pp\ collisions. 

TA correlations can also be constructed for restricted azimuth intervals relative to the trigger momentum. Of special interest is the transverse region (TR), an azimuth interval bracketing $\pi/2$ relative to the trigger direction. Underlying event (UE) analysis assumes that a triggered dijet is confined within jet cones at 0 and $\pi$ and should not contribute to the TR. The TA hard component extracted from the TR may challenge that assumption.

The TA TCM established in this study can be applied to both \pp\ and \aa\ data. The TA hard component may provide new insights into jet production from nuclear collisions, especially modified jet formation in more-central \aa\ collisions. Application to Monte Carlo data may test basic assumptions invoked by QCD models, including the structure of the scattered-parton spectrum and the frequency of multiple parton interactions (MPI).

This work was supported in part by the Office of Science of the U.S.\ DOE under grant DE-FG03-97ER41020. 

\begin{appendix}

\section{Marginal Constraints} \label{marg}

According to the two-component model of high-energy nuclear collisions hadron production arises mainly from projectile-nucleon dissociation (soft) or large-angle-scattered parton fragmentation (hard). The TCM for SP spectra assumes that all final-state hadrons are sampled from single-particle model spectra for soft or hard events, and the latter from soft or hard spectrum components. The TCM 2D TA distribution must project to trigger and associated 1D marginal spectra that are consistent with SP spectrum structure. Model construction is based on $F_x(y_t,n_{ch})$, the unit-normal SP spectrum model for event type $x$ corresponding to spectrum data from multiplicity class $n_{ch}$. The goal is TCM reference $F_x(y_{ta},y_{tt},n_{ch})$, the most general model for trigger-associated correlations consistent with marginal constraints and a factorization assumption representing minimal TA correlations.

\subsection{Marginal spectra}

$T_x(y_{tt},n_{ch})$ is the per-event marginal trigger-particle spectrum integrating to  one trigger particle per event. The total charged-particle number within the angular acceptance is $n_{ch} = \Delta \eta\, dn_{ch}/d\eta$. The trigger spectrum is derived from the unit-normal SP spectrum $F_x(y_t,n_{ch})$ by $T_x(y_{tt},n_{ch}) = G_x(y_{tt},n_{ch})\, n_{ch} F_x(y_{tt},n_{ch})$, where $G_x(y_{tt},n_{ch})$ is the void probability that no samples appear above $y_{tt}$ [defined by Eqs.~(\ref{nsum}) and (\ref{g0yt})]. 
Trigger particles may appear on the main diagonal on $(y_{ta},y_{tt})$ as a self-pair contribution to TA correlations.  The trigger spectrum then has the same form in projections onto $y_{ta}$ and $y_{tt}$. Self pairs are excluded from the present analysis. 

The per-event marginal associated-particle spectrum is denoted by $A_x(y_{ta})$. The SP spectrum for each $n_{ch}$ class is the sum of trigger and associated spectra
\bea
n_{ch} F_x(y_{ta}) &=& T_x(y_{ta}) + (n_{ch} - 1)A_x(y_{ta}),
\eea
leading to an expression for associated-particle spectra
\bea \label{axdef}
(n_{ch}-1) A_x(y_{ta}) &=& [1 - G_x(y_{ta})]\, n_{ch} F_x(y_{ta}).
\eea

\subsection{TCM marginal constraints}

2D TA correlations can be factorized according to Bayes' theorem as $F_x(y_{ta},y_{tt}) = T_x(y_{tt})\, A_x(y_{ta}:y_{tt})$. We seek the most general form for conditional spectra $ A_x(y_{ta}:y_{tt})$ consistent with imposed constraints. The definition $A_x(y_{ta}:y_{tt}) =  D_x(y_{tt}) E_x(y_{ta})F_x(y_{ta})$ (for $y_{ta} < y_{tt}$) assumes SP distributions $F_x(y_{ta})$ as the initial case. $O(1)$ weight functions $D_x(y_{tt})$ and $E_x(y_{ta})$ (with initial values 1) represent the effect of marginal constraints as described below.

The first constraint is defined by projecting $F_x(y_{ta},y_{tt})$ onto $y_{tt}$ to obtain $T_x(y_{tt})$.  Assuming initial values $E_x = 1$ and canceling common factor $T_x(y_{tt})$ gives
\bea \label{dx}
D_x(y_{tt}) \int_0^{y_{tt}} dy_{ta} y_{ta}  F_x(y_{ta})  &=& 1,
\eea
which defines weight function $D_x(y_{tt}) \geq 1$. That constraint is equivalent to requiring that any event in multiplicity class $n_{ch}$ contains $n_{ch}-1$ associated particles.

The second constraint is defined by projecting $F_x(y_{ta},y_{tt})$ onto $y_{ta}$ to obtain marginal spectrum $A_x(y_{ta})$, effectively the trigger-weighted average of conditional associated spectra. The uncorrected projection onto $y_{ta}$ is 
\bea \label{margint}
I_x(y_{ta}) 
&=& F_x(y_{ta}) \int_{y_{ta}}^{\infty} dy_{tt} y_{tt} D_x(y_{tt}) T_x(y_{tt}).
\eea
Combining Eqs.~(\ref{axdef}) and (\ref{margint}) in the form $A_x = E_x I_x$ given $D_x(y_{tt})$ from Eq.~(\ref{dx}) then determines weight functions $E_x(y_{ta}) \approx 1$.  The combined weight functions define the primed associated-particle spectra referred to in the text as $A_s = S_0'' = D_s E_s S_0$ and $A_h = p_s' S_0' + p_h' H_0' = D_h E_h (p_s' S_0 + p_h' H_0)$ (for $y_{ta} < y_{tt}$).

The marginal constraints simplify for certain limiting cases. $D_x(y_{tt})$ is $\gg 1$ for $y_{tt}$ small and $\approx 1$ for $y_{tt}$ large. 
$G_x(y_{t}) \rightarrow 1$ for $y_{t}$ large and $1 - G_x(y_{t}) \rightarrow 1$ for $y_{t}$ small. 
For $y_{ta}$ large (and therefore $y_{tt}$ large) Eq.~(\ref{margint}) becomes
\bea \label{runint}
I_x(y_{ta}) 
&\approx &  F_x(y_{ta})  D_{x0} \int_{y_{ta}}^{\infty} dy_{tt} y_{tt} n_{ch}  F_x(y_{tt}) \\ \nonumber
&\approx & -\ln[G_x(y_{ta})] D_{x0} F_x(y_{ta})
\eea
where $D_{x0} \approx 1$ is the limiting value of $D_x$ for large $y_{tt}$. The first line follows from $G_x \approx 1$, and the second line follows from $n_{x\Sigma} = -\ln(G_x) \approx 1 - G_x$ defined by Eq.~(\ref{nsum}). That result is consistent with $A_x(y_{ta})$ from Eq.~(\ref{axdef}) for $y_{ta}$ large.
 For $y_{ta}$ small Eq.~(\ref{margint}) becomes
\bea 
I_x(y_{ta}) &\approx&  F_x(y_{ta})  \int_{y_{ta}}^{\infty} dy_{tt} y_{tt}D_x(y_{tt})  T_x(y_{tt})  \\ \nonumber
&\approx& \langle D_x(y_{ta})\rangle F_x(y_{ta})
\eea
also consistent with Eq.~(\ref{axdef}) for $y_{ta}$ small.
%
For the following we use simulation data to compare marginal associated-particle spectra $A_x(y_{ta})$ as defined by Eq.~(\ref{axdef}) with projection integrals $I_x(y_{ta})$ defined by Eq.~(\ref{margint}), separately for soft and hard events and for four $n_{ch}$ classes.

\subsection{Simulation results}

 \begin{figure}[h]
  \includegraphics[width=3.3in]{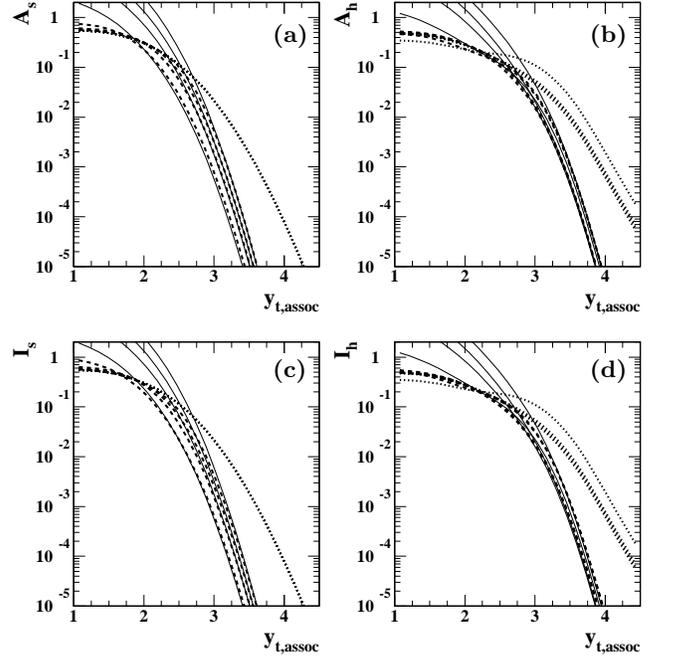}
\put(-20,105) {\bf (d)}
 \put(-20,232) {\bf (b)}
 \put(-140,105) {\bf (c)}
 \put(-140,232) {\bf (a)}
\caption{\label{specc3}
a), (b): Predicted marginal associated-particle spectra $A_x(y_{ta},n_{ch})$ (dashed curves) as defined by Eq.~(\ref{axdef}). Limiting cases are represented by $-\ln[G_x(y_{ta},n_{ch})] F_x(y_{ta},n_{ch})$ (thin solid curves) for larger $y_{ta}$ and SP model functions $F_x(y_{ta},n_{ch})$ (dotted curves) for smaller $y_{ta}$. 
(c), (d):  Running integrals $I_x(y_{ta},n_{ch})$ (dashed curves) as defined by Eq.~(\ref{margint}). The limiting cases (thin solid and dotted curves) are the same as for the upper panels.
 }  
 \end{figure}

Figure~\ref{specc3} (a), (b) shows marginal associated-particle spectra $A_x(y_{ta},n_{ch})$ 
(dashed curves) defined by Eq.~(\ref{axdef}) compared to limiting cases $F_x(y_{ta})$ (for smaller $y_{ta}$, dotted curves) and $-\ln[G_x(y_{ta})]F_x(y_{ta})$ (for larger $y_{ta}$, thin solid curves).
Figure~\ref{specc3} (c), (d) shows running integrals $I_x(y_{ta})$ for four $n_{ch}$ classes (dashed curves) defined by Eq.~(\ref{margint}) compared to the same limiting  cases. The lowest dashed curves are for the lowest $n_{ch}$ class.
The $F_s$ for all soft events are the same (by construction), whereas the  $F_h$ for hard events depend on $n_{ch}$.  Because all hard events include at least one jet those events with the smallest $n_{ch}$ have the hardest spectra (uppermost dotted curves in right panels). It is the {\em fraction of hard events} in any $n_{ch}$ class that increases with $n_{ch}$, thus making the observed ensemble-averaged spectra harder.

 \begin{figure}[h]
  \includegraphics[width=3.3in]{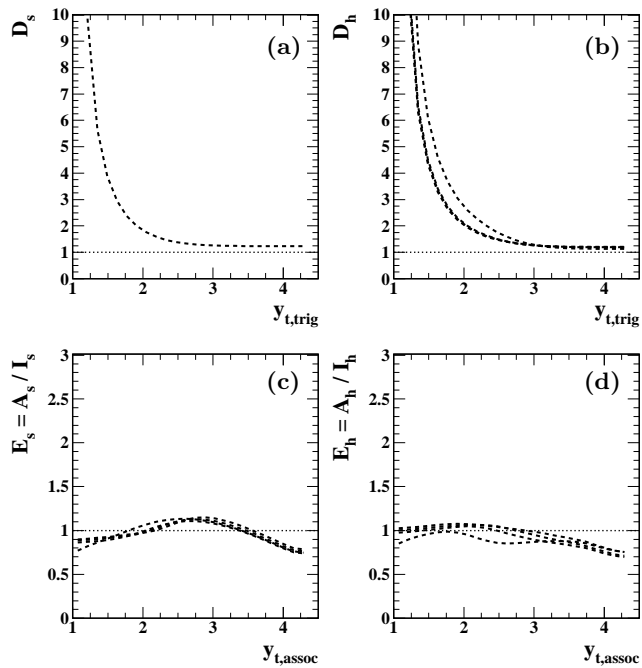}
\put(-22,105) {\bf (d)}
 \put(-22,232) {\bf (b)}
 \put(-142,105) {\bf (c)}
 \put(-142,232) {\bf (a)}
\caption{\label{specc4}
(a), (b): Weight functions $D_x(y_{tt})$ defined by Eq.~(\ref{dx}). The separate curve in the right panel corresponds to the lowest $n_{ch}$ bin.
(c), (d): Weight functions $E_x(y_{tt})$ defined by the vertical axis label and Eqs.~(\ref{axdef}) and (\ref{margint}).
 }  
 \end{figure}

Figure~\ref{specc4} (a), (b) shows $D_x(y_{tt})$ inferred from Eq.~(\ref{dx}). The limiting values for larger $y_{tt}$ denoted by $D_{x0}$ reflect the effects of integration over incomplete $y_t$ acceptance and possible detector inefficiency (for real data).  The data integration on $y_{tt}$ or $y_{ta}$ extends over $[1,4.5]$, not [$0,\infty$] and thus excludes a significant fraction $O(10\%)$ of the SP spectrum soft component.
Figure~\ref{specc4} (c), (d) shows $E_x(y_{ta})$  inferred from the combination of Eqs.~(\ref{axdef}) and~(\ref{margint}). The deviations from unity are small. The exceptional curves in the right panels correspond to the smallest $n_{ch}$ class and therefore the hardest spectrum.

Because both $E_x$ factors are close to unity for smaller $y_{ta}$ they would not significantly change the $D_x$ inferred from Eq.~(\ref{dx}). A single normalization iteration is then sufficient. 
Within the interval $y_{ta} < 2.5$ where the soft-component model is relevant for isolation of a data hard component $E_s $ remains within 10\% of unity.
The hard-component weight functions $E_h$ for all but the lowest $n_{ch}$ class do not deviate more than 15\% from unity, suggesting that corrections $H' \rightarrow H$ for marginal distortions may not be required. These results confirm the accuracy of numerical integration over several decades and reveal that the TA TCM is self-consistent to about 10\%.

\section{TCM model functions}  \label{tcmmodelfunc} 

We summarize the TCM single-particle spectrum model functions that provide the basis for this 2D trigger-associated analysis. The soft-component model is a limiting case of measured spectra. The hard-component model is quantitatively consistent with a pQCD parton spectrum folded with measured jet fragmentation functions to describe a parton fragment distribution~\cite{fragevo}.

\subsection{SP spectrum soft and hard model functions}

The unit-integral functions for the two-component model (TCM) of $m_t$ or $y_t$ spectra used in this analysis are defined in Refs.~\cite{ppprd,hardspec}. For 200 GeV \pp\ collisions the soft-component model (L\'evy distribution on $m_t$) is
\bea
S_0(y_t) = \frac{20.6}{[1 + (m_t - m_\pi)/{nT}]^n}
\eea
with $m_t = m_\pi \cosh(y_t)$, $n = 12.8$ and $T = 0.145$ GeV. The Gaussian form of the hard-component model is
\bea
H_0(y_t) = 0.33 \exp\{-(y_t - y_{t0})^2/2\sigma_{y_t}^2\}
\eea
on $y_t$, with $y_{t0} = 2.67$ ($p_t \approx 1$ GeV/c) and $\sigma_{y_t} = 0.445$. The coefficients (determined by the unit-integral condition) depend on the specific model parameters. 

\subsection{Constructing a power-law tail} \label{power}

In Ref.~\cite{hardspec} hard-component model $H_0(y_t)$ is generalized to a Gaussian with added power-law tail to accommodate the underlying parton energy spectrum.  The power-law trend $\exp(-q\, y_t)$ appears as a straight line when $\ln(H_0)$ is plotted vs $\ln p_t$ or vs $y_t \sim \ln(p_t)$. Parameter $q$ is the power-law exponent on $y_t$ (different from that on $p_t$). $H_0(y_t)$ is constructed as a Gaussian function with power-law tail as follows. For given Gaussian parameters the Gaussian trend transitions to the power-law trend (slopes equal) at $y_t- y_{t0} = q \sigma_{y_t}^2 $ where the exponent of $H_0(y_t)$ is $q^2 \sigma_{y_t}^2/2$. The exponent function below that point is $(y_t - y_{t0})^2/2\sigma_{y_t}^2$ and above that point is $q(y_t - y_{t0}) - q^2 \sigma_{y_t}^2/2$. The required function $H_0(y_t)$ is obtained by exponentiating those functions within the specified $y_t$ intervals. For the TCM parameters used in this analysis (including $q \approx 5.5$) the transition to power-law tail occurs near $y_t = 3.75$ (see Fig.~\ref{ppspectra}, right). 

In Ref.~\cite{fragevo} a pQCD-calculated minimum-bias fragment distribution derived from measured jet fragmentation functions corresponds well with the Gaussian+tail model of the spectrum hard component except below 0.5 GeV/c ($y_t = 2$) where the Gaussian model drops below the pQCD calculation. Within that same low-$p_t$ interval systematic uncertainties in the inferred spectrum hard component and pQCD prediction are relatively large. 

\end{appendix}



\begin{thebibliography}{99}

\bibitem{pptheory}  T.~A.~Trainor,
Phys.\ Rev.\ D {\bf 87}, 054005 (2013).

\bibitem{fragevo}    T.~A.~Trainor,
  Phys.\ Rev.\  C {\bf 80}, 044901 (2009).

\bibitem{anomalous}  G.\ Agakishiev, {\it et al.} (STAR Collaboration),
  Phys.\ Rev.\ C {\bf 86}, 064902 (2012).

\bibitem{opal}  M.~Z. Akrawy {et al.}  (OPAL Collaboration)
  {Phys. Lett.} B, {\bf 247}, 617 (1990).

\bibitem{tasso} W.~Braunschweig {\it et al.}  (TASSO Collaboration),
Z.\ Phys.\ C {\bf 47}, 187 (1990). 

\bibitem{eeprd}   T.~A.~Trainor and D.~T.~Kettler,
  Phys.\ Rev.\ D {\bf 74}, 034012 (2006).

\bibitem{ppprd} J.~Adams {\it et al.}  (STAR Collaboration),
  Phys.\ Rev.\  D {\bf 74}, 032006 (2006).

\bibitem{cdf}  T.~Affolder {\it et al.}  (CDF Collaboration),
  Phys.\ Rev.\ D {\bf 65}, 092002 (2002).

\bibitem{ua1} C.~Albajar {\it et al.}  (UA1 Collaboration),
  Nucl.\ Phys.\  B {\bf 309}, 405 (1988).

\bibitem{starhipt}  M.~M.~Aggarwal {\it et al.}  (STAR Collaboration),
  Phys.\ Rev.\ C {\bf 82}, 024912 (2010).

\bibitem{tzyam} T.~A.~Trainor,
  Phys.\ Rev.\  C {\bf 81}, 014905 (2010).

\bibitem{axialci}   J.~Adams {\it et al.}  (STAR Collaboration),
  Phys.\ Rev.\  C {\bf 73}, 064907 (2006).

\bibitem{porter2} R.~J.~Porter and T.~A.~Trainor  (STAR Collaboration),
  J.\ Phys.\ Conf.\ Ser.\  {\bf 27}, 98 (2005).

\bibitem{porter3}  R.~J.~Porter and T.~A.~Trainor  (STAR Collaboration),
  PoS C {\bf FRNC2006}, 004 (2006).

\bibitem{hardspec}  T.~A.~Trainor,
  Int.\ J.\ Mod.\ Phys.\  E {\bf 17}, 1499 (2008).

\bibitem{rick} R.~Field,
  Acta Phys.\ Polon.\ B {\bf 42}, 2631 (2011).

\bibitem{2004}   J.~Adams {\it et al.}  (STAR Collaboration),
  Phys.\ Rev.\  C {\bf 72}, 014904 (2005).

\bibitem{multipoles}  T.~A.~Trainor,
  J.\ Phys.\ G {\bf 40}, 055104 (2013).

\bibitem{gluequad}  T.~A.~Trainor,
  Mod.\ Phys.\ Lett.\  A {\bf 23}, 569 (2008).

\bibitem{starbulk}   M.~M.~Aggarwal {\it et al.}  (STAR Collaboration),
  Phys.\ Rev.\ C {\bf 83}, 034910 (2011).

\bibitem{jetspec}   T.~A.~Trainor and D.~T.~Kettler,
  Phys.\ Rev.\ C {\bf 83}, 034903 (2011).

\bibitem{axialcd} J.~Adams {\it et al.}  (STAR Collaboration),
  Phys.\ Lett.\  B {\bf 634}, 347 (2006).

\bibitem{inverse} T.~A.~Trainor, R.~J.~Porter and D.~J.~Prindle,
  J.\ Phys.\ G {\bf 31}, 809 (2005).

\bibitem{kn}  D.~Kharzeev and M.~Nardi,
  Phys.\ Lett.\ B {\bf 507}, 121 (2001).

\bibitem{nov2} T.~A.~Trainor, D.~T.~Kettler, D.~J.~Prindle and R.~L.~Ray,
  arXiv:1302.0300.

\bibitem{herwig}  M.~Bahr, S.~Gieseke, M.~A.~Gigg, D.~Grellscheid, K.~Hamilton, O.~Latunde-Dada, S.~Platzer and P.~Richardson {\it et al.},
  Eur.\ Phys.\ J.\ C {\bf 58}, 639 (2008).

\bibitem{herwig2} M.~Bahr, J.~M.~Butterworth and M.~H.~Seymour,
  JHEP {\bf 0901}, 065 (2009).

\bibitem{hijing} X.-N. Wang,  Phys. Rev. D {\bf 46}, R1900 (1992); 
X.-N.~Wang and M.~Gyulassy,
  Phys.\ Rev.\  D {\bf 44}, 3501 (1991).

\bibitem{pythia}  T. Sj\"ostrand and M. van Zijl, Phys. Rev. D {\bf 36}, 2019 (1987);
T.~Sj\"ostrand,
Comput.\ Phys.\ Commun.\  {\bf 82}, 74 (1994); 
T.~Sj\"ostrand, L.~L\"onnblad, S.~Mrenna and P.~Skands,
hep-ph/0308153.

\bibitem{liwang} S.-y.~Li and X.-N.~Wang,
  Phys.\ Lett.\ B {\bf 527}, 85 (2002).

\bibitem{review} T.~A.~Trainor,
  arXiv:1303.4774. 

\bibitem{jetmult}  I.~M.~Dremin and J.~W.~Gary,
  Phys.\ Rept.\  {\bf 349}, 301 (2001).

\bibitem{cleo} M.~S.~Alam {\it et al.}  (CLEO Collaboration),
  Phys.\ Rev.\ D {\bf 56}, 17 (1997). 

\bibitem{starraa} B.~I.~Abelev {\it et al.}  (STAR Collaboration),
  Phys.\ Rev.\ C {\bf 81}, 054907 (2010).

\bibitem{starzyam}  M.~M.~Aggarwal {\it et al.}  (STAR Collaboration),
  Phys.\ Rev.\ C {\bf 82}, 024912 (2010).

\bibitem{cdfue}   T.~Affolder {\it et al.}  (CDF Collaboration),
  Phys.\ Rev.\ D {\bf 65}, 092002 (2002).

\bibitem{cmsue}  V.~Khachatryan {\it et al.}  (CMS Collaboration),
  Eur.\ Phys.\ J.\ C {\bf 70}, 555 (2010).

\end{thebibliography}
\end{document}